\documentclass[aps,pra, twocolumn,superscriptaddress,longbibliography,nofootinbib]{revtex4-2}

\usepackage{graphicx}

\usepackage{derivative}
\usepackage{xcolor}
\usepackage[margin=0.68in]{geometry}
\usepackage{siunitx}
\usepackage{upgreek}
\usepackage{soul}
\usepackage{subfiles}
\usepackage{color}
\usepackage{amssymb}
\usepackage{amsmath, scalerel, breqn}
\usepackage{bbm}
\usepackage{physics}
\usepackage{bbold}
\usepackage{ulem}

\definecolor{dark-red}{rgb}{0.4,0.15,0.15}
\definecolor{dark-blue}{rgb}{0.15,0.15,0.4}
\definecolor{medium-blue}{rgb}{0,0,0.5}
\usepackage{hyperref}
\hypersetup{
    colorlinks, linkcolor={dark-red},
    citecolor={dark-blue}, urlcolor={medium-blue}
}

\usepackage{xpatch}
\makeatletter
\patchcmd{\@ssect@ltx}
    {\addcontentsline{toc}{#1}{\protect\numberline{}#8}}
    {}
    {}
    {}
\makeatother

\definecolor{green}{rgb}{0.08,0.7,0.05}

\definecolor{xm}{rgb}{1,0,0}

\usepackage{nicefrac}
\usepackage{hyperref}
\usepackage{xfp}

\usepackage{etoolbox}

\newcounter{manualbibstart}
\makeatletter
\apptocmd{\thebibliography}{%
  \setcounter{NAT@ctr}{\value{manualbibstart}}%
  \addtocounter{NAT@ctr}{-1}%
}{}{}
\makeatother

\begin{document}

\setcitestyle{super}

\title{Thermalization and Criticality on an Analog-Digital Quantum Simulator} 

\maketitle
\onecolumngrid
\vspace{-1cm}

\begin{flushleft}
   
    \newcommand\blfootnote[1]{%
          \begingroup
          \renewcommand\thefootnote{}\footnote{#1}%
          \addtocounter{footnote}{-1}%
          \endgroup
        }

    \renewcommand{\author}[2]{#1\textsuperscript{\textrm{\scriptsize #2}}}
    \renewcommand{\affiliation}[2]{\textsuperscript{\textrm{\scriptsize #1} #2} \\}
    \newcommand{\corrauthora}[2]{#1$^{\textrm{\scriptsize #2}, \hyperlink{corra}{\ddagger}}$}
    \newcommand{\corrauthorc}[2]{#1$^{\textrm{\scriptsize #2}, \hyperlink{corra}{\ddagger},\hyperlink{corrb}{\mathsection}}$}
    \newcommand{\corrauthorb}[2]{#1$^{\textrm{\scriptsize #2}, \hyperlink{corrb}{\mathsection}}$}
    \newcommand{\xGoogle}{\affiliation{$^{1}$}{Google Research, Mountain View, CA, USA}}
\newcommand{\xGeneva}{\affiliation{$^{2}$}{Department of Theoretical Physics, University of Geneva, Quai Ernest-Ansermet 24, 1205 Geneva, Switzerland}}

\newcommand{\xLTCP}{\affiliation{$^{13}$}{Laboratory for Theoretical and Computational Physics, Paul Scherrer Institute, 5232 Villigen, Switzerland}}

\newcommand{\xEPFL}{\affiliation{$^{14}$}{Institute of Physics, Ecole Polytechnique F\'{e}d\'{e}rale de Lausanne (EPFL), CH-1015 Lausanne, Switzerland}}

\newcommand{\xUMD}{\affiliation{$^{3}$}{Joint Quantum Institute and Joint Center for Quantum Information and Computer Science, NIST/University of Maryland, College Park, MD, USA}}

\newcommand{\xPrinceton}{\affiliation{$^{15}$}{Department of Physics, Princeton University, Princeton, NJ, USA}}

\newcommand{\xUMass}{\affiliation{$^{8}$}{Department of Electrical and Computer Engineering, University of Massachusetts, Amherst, MA, USA}}

\newcommand{\xUConn}{\affiliation{$^{5}$}{Department of Physics, University of Connecticut, Storrs, CT, USA}}

\newcommand{\xAuburn}{\affiliation{$^{9}$}{Department of Electrical and Computer Engineering, Auburn University, Auburn, AL, USA}}

\newcommand{\xUCR}{\affiliation{$^{11}$}{Department of Electrical and Computer Engineering, University of California, Riverside, CA, USA}}

\newcommand{\xHarvard}{\affiliation{$^{12}$}{Department of Chemistry and Chemical Biology, Harvard University, MA, USA}}

\newcommand{\xSydney}{\affiliation{$^{10}$}{QSI, Faculty of Engineering \& Information Technology, University of Technology Sydney, NSW, Australia}}

\newcommand{\xIMM}{\affiliation{$^{4}$}{Institute of Molecules and Materials, Radboud University, Nijmegen, Netherlands}}

\newcommand{\xCNRS}{\affiliation{$^{7}$}{Universit\'e Grenoble Alpes, CNRS, LPMMC, Grenoble,
France}}
\newcommand{\xCaltech}{\affiliation{$^{6}$}{Institute for Quantum Information and Matter and Walter Burke Institute for Theoretical Physics, Caltech, Pasadena, CA, USA}}

\begin{footnotesize}

\newcommand{\Google}{1}
\newcommand{\Geneva}{2}
\newcommand{\UMD}{3}
\newcommand{\IMM}{4}
\newcommand{\UConn}{5}
\newcommand{\Caltech}{6}
\newcommand{\CNRS}{7}
\newcommand{\UMass}{8}
\newcommand{\AU}{9}
\newcommand{\Sydney}{10}
\newcommand{\UCR}{11}
\newcommand{\Harvard}{12}
\newcommand{\LTCP}{13}
\newcommand{\EPFL}{14}
\newcommand{\Princeton}{15}

\corrauthorc{T. I.~Andersen}{\Google},
\corrauthora{N. Astrakhantsev}{\Google},
\author{A. H. Karamlou}{\Google},
\author{J. Berndtsson}{\Google},
\author{J. Motruk}{\Geneva},
\author{A. Szasz}{\Google},
\author{J.~A.~Gross}{\Google},
\author{A. Schuckert}{\UMD},
\author{T. Westerhout}{\IMM},
\author{Y. Zhang}{\Google},
\author{E. Forati}{\Google},
\author{D. Rossi}{\Geneva},
% ...
\corrauthorb{D.~A. Abanin}{\Google,\!\Princeton},
\corrauthorb{X. Mi}{\Google}

\begin{center}
\textit{
\xGoogle
\xGeneva
\xUMD
\xIMM
\xUConn
\xCaltech
\xCNRS
\xUMass
\xAuburn
\xSydney
\xUCR
\xHarvard
\xLTCP
\xEPFL
\xPrinceton}
\end{center}

\end{footnotesize}
\blfootnote{{\hypertarget{corra}{${}^\ddagger$} These authors contributed equally to this work.}\\
{\hypertarget{corrb}{${}^\mathsection$} Corresponding author: trondiandersen@google.com}\\
{\hypertarget{corrb}{${}^\mathsection$} Corresponding author: abanin@google.com}\\
{\hypertarget{corrb}{${}^\mathsection$} Corresponding author: mixiao@google.com}\\}
\end{flushleft}

\vspace{-1cm}

{\bf Understanding how interacting particles approach thermal equilibrium is a major challenge of quantum simulators~\cite{Altman_PRXQ_2021,AbaninRMP2019}. Unlocking the full potential of such systems toward this goal requires flexible initial state preparation, precise time evolution, and extensive probes for final state characterization. We present a quantum simulator comprising 69 superconducting qubits which supports both universal quantum gates and high-fidelity analog evolution, with performance beyond the reach of classical simulation in cross-entropy benchmarking experiments. Emulating a two-dimensional (2D) XY quantum magnet, we leverage a wide range of measurement techniques to study quantum states after ramps from an antiferromagnetic initial state. We observe signatures of the classical Kosterlitz-Thouless phase transition~\cite{KosterlitzThouless1973}, as well as strong deviations from Kibble-Zurek scaling predictions~\cite{DelCampoZurekReview2014} attributed to the interplay between quantum and classical coarsening of the correlated domains~\cite{SamajdarArxiv2024}. This interpretation is corroborated by injecting variable energy density into the initial state, which enables studying the effects of the eigenstate thermalization hypothesis (ETH)~\cite{DeutschETH,SrednickiETH,Polkovnikov-rev} in targeted parts of the eigenspectrum. Finally, we digitally prepare the system in pairwise-entangled dimer states and image the transport of energy and vorticity during thermalization. These results establish the efficacy of superconducting analog-digital quantum processors for preparing states across many-body spectra and unveiling their thermalization dynamics.}

\clearpage
\twocolumngrid
The advent of quantum simulators in various platforms~\cite{daley2022practical,BlochColdAtoms,Blatt12,HouckNaturePhysicsReview2012,carusotto2020photonic,BrowaeysNaturePhysicsReview2020,king2024computational} has opened a powerful experimental avenue toward answering the theoretical question of thermalization~\cite{DeutschETH,SrednickiETH}, which seeks to reconcile the unitarity of quantum evolution with the emergence of statistical mechanics in constituent subsystems. A particularly interesting setting for studying thermalization is that in which a quantum system is swept through a critical point~\cite{Dziarmaga2005,ZurekZollerPRL2005,PolkovnikovPRB2005}, since varying the sweep rate can allow for accessing dramatically different paths through phase space and correspondingly distinct coarsening behavior. In systems with complex phase diagrams involving multiple phases, such effects have been theoretically predicted to cause deviations~\cite{SamajdarArxiv2024} from the celebrated Kibble-Zurek (KZ) mechanism, which states that the correlation length $\xi$ of the final state follows a universal power-law scaling with the ramp time $t_{\text{r}}$~\cite{DelCampoZurekReview2014,keesling2019quantum,ebadi2021quantum}.

\begin{figure}[!htbp]
    \centering \includegraphics[width=1.0\columnwidth]{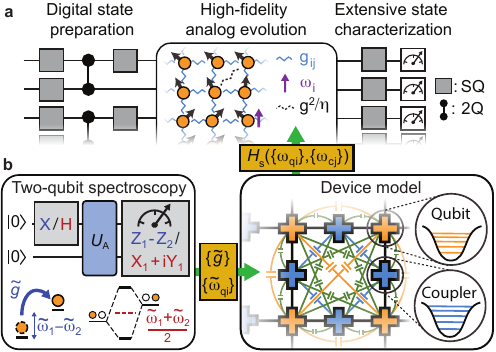}
    \caption{{\bf Analog-digital simulation with high-precision calibration. a,} Our platform combines analog evolution with digital gates for extensive state preparation and characterization. {\bf b,} Schematic of new scalable analog calibration scheme. Swap (blue) and single-photon (red) spectroscopy is used to extract dressed coupling rates ($\{\tilde{g}\}$) and qubit frequencies ($\{\tilde{\omega}_{qi}\}$) of 2-qubit analog evolution ($U_A$), which are converted to bare qubit and coupler frequencies ($\{\omega_{qi}\},\{\omega_{cj}\}$) through detailed device modeling. The bare frequencies allow for establishing the device Hamiltonian of the full system, which is finally projected to a spin-Hamiltonian, $H_{\text{s}}$.}
    \label{fig:1}
\end{figure}
While tremendous technical advancements in quantum simulators have enabled the observation of a wealth of thermalization-related phenomena~\cite{SchmiedmeyerScience2015GGE,SchreiberMBLScience15,KaufmanScience2016, neill_2016, roushan_2017, braumuller_2022, zhou_2022, zhang_2023}, the analog nature of these systems has also imposed constraints on the experimental versatility. Studying thermalization dynamics necessitates state characterization beyond density-density correlations and preparation of initial states across the entire eigenspectrum, both of which are difficult without universal quantum control~\cite{karamlou_2023}. While digital quantum processors are in principle suitable for such tasks, implementing Hamiltonian evolution requires a high number of digital gates, making large-scale Hamiltonian simulation infeasible under current gate errors.
\begin{figure*}[t!]
    \centering
    \includegraphics[width=1.0\textwidth]{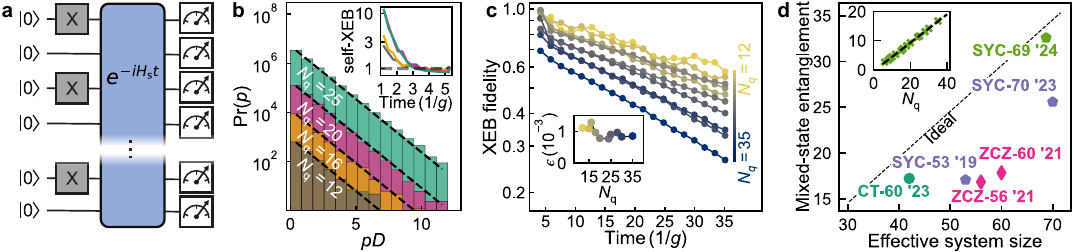}
    \caption{{\bf Fast thermalization dynamics and beyond-classical capabilities in the high-temperature regime.\newline a,} Schematic representation of the experiment: $N_{\mathrm{q}}$ qubits are initialized in a half-filling state, evolved under a Hamiltonian $H_{\text{s}}$ over time $t$ with $4$ instances of disorder in $\{\omega_i\}$, and finally measured in the $Z$-basis. {\bf b,} Distribution $\text{Pr}(p)$ of bitstring probabilities $p$ from experiment (colored bars) and ideal PT distribution $\text{Pr}(p) = D e^{-p D}$ (dashed lines). Inset: Convergence of the self-XEB with time. {\bf c,} Time-dependent XEB fidelity for system sizes up to $N_{\mathrm{q}} = 35$. Inset: System size dependence of $\epsilon$ (error per qubit per evolution time of $1/g$) from exponential fits. {\bf d,} Mixed-state entanglement proxy, $\mathcal{E}_{\text{P}}$, obtained in this and previous works, plotted against the effective system size $N^{\text{eff}}_{\text{q}}$ (with respect to entanglement of a fully chaotic state; SI) of the respective platforms. Blue pentagons: Sycamore processor in the digital regime~\cite{AruteNature19,morvan2023phase}; diamonds: Zuchongzhi processor~\cite{ZuchongzhiPRL21,ZuchongZhiSciBull2021}; circle: neutral atom analog simulator~\cite{ShawNature2024}; green pentagon: present experiment. $N^{\text{eff}}_{\text{q}}$ is equal to the actual $N_{\text{q}}$ in the digital experiments, while analog platforms are subject to $U(1)$ conservation (this work) or constraints from Rydberg blockade~\cite{ShawNature2024}. Inset: $\mathcal{E}_{\text{P}}$ as a function of $N_{\mathrm{q}}$ computed from experimental data, including the linear fit used for extrapolation to 69 qubits.}
    \label{fig:2}
\end{figure*}

In this work, we present a hybrid analog-digital~\cite{bluvstein2022quantum,Lamata2018} quantum simulator comprising 69 superconducting transmon qubits connected by tunable couplers in a 2D lattice~(Fig.~\ref{fig:1}a). The quantum simulator supports universal entangling gates with pairwise interaction between qubits, and high-fidelity analog simulation of a $U(1)$ symmetric spin-Hamiltonian when all couplers are activated at once. The low analog evolution error, which was previously difficult to achieve with transmon qubits due to correlated cross-talk effects, is enabled by a new scalable calibration scheme (Fig.~\ref{fig:1}b). Using cross-entropy benchmarking (XEB)~\cite{boixo_2018}, we demonstrate analog performance that exceeds the simulation capacity of known classical algorithms at the full system size.

Leveraging these capabilities, we prepare and characterize states of a 2D XY-magnet with broadly tunable energy density, allowing us to study the interplay between quantum and classical critical coarsening in the rich phase diagram of our system. Specifically, we observe finite-size signatures of the Kosterlitz-Thouless (KT) topological phase transition --- including the emergence of algebraically decaying correlations with exponent near the expected universal value of $\frac{1}{4}$ --- and demonstrate a resultant breakdown of the KZ mechanism. Our study takes advantage of measurements that go beyond standard two-point correlators, to characterize entanglement entropy for subsystems up to 12 qubits, multi-qubit vortex correlators, and energy fluctuations. We also leverage our hybrid analog-digital scheme (Fig.~\ref{fig:1}a) to prepare entangled initial states, allowing us to tailor the spatial distribution of energy density and vorticity, and investigate the subsequent thermalization dynamics and energy transport.

 Operating coupled transmons as a high-fidelity analog quantum simulator requires precise knowledge of the many-body Hamiltonian $H_{\text{s}}$, which depends on the ``bare'' qubit and coupler frequencies, $\{\omega_{qi}\}$ and $\{\omega_{cj}\}$. However, experimental calibration is only capable of resolving ``dressed`` frequencies which --- unlike the bare frequencies --- change from local (isolated) calibrations to full-scale experiments due to hybridization with neighboring qubits and couplers. Given this difficulty, past studies either suffered from large errors or resorted to multi-parameter learning protocols that are difficult to scale up~\cite{neill_2016,roushan_2017}. 
 
In this work, we have developed a scalable calibration protocol that achieves low error by explicitly calibrating the bare frequencies. As illustrated in Fig.~\ref{fig:1}b, the protocol begins with two-qubit calibration measurements (single-photon and swap spectroscopy) to determine the effective coupling $\tilde{g}$ and dressed qubit frequencies $\{\tilde{\omega}_{qi}\}$ of every qubit pair. Next, we employ extensive modeling of the underlying device physics to convert the dressed quantities to the bare frequencies $\{\omega_{qi}\}$,$\{\omega_{cj}\}$. Finally, a projection technique is applied to approximate our high-dimension device Hamiltonian, $H_d(\{\omega_{qi}\},\{\omega_{cj}\})$, into a spin-Hamiltonian, $H_{\text{s}}$:
\begin{align}\label{eq:H}
&H_{\text{s}}=\sum_i \omega_i n_i+\sum_{\langle i,\,j\rangle} g_{ij}(X_iX_j+Y_iY_j)/2+\mathcal{O}(g^2/\eta) 
\end{align}
where $\omega_i$ and $|g_{ij}|\,\approx g$ are tunable on-site potentials and nearest-neighbor couplings, respectively. The latter is significantly smaller than the qubit anharmonicity $\eta\gg g$. This restricts the photon occupation numbers to $n_i=0,1$, and $X_i,Y_i$ are Pauli operators acting in this subspace. Using local calibrations and accurate modeling of the underlying device physics makes our approach more scalable than fitting a large number of parameters. The Hamiltonian in Eq.\,\eqref{eq:H} is in the universality class of an XY model with on-site $z$-fields. A natural consequence of the hybridization in our system is that $H_{\text{s}}$ contains not only nearest-neighbor hopping, but also density-density interactions and next-nearest-neighbor terms, which scale as $g^2/\eta$ and are typically 5-10 times smaller than $g$.

\begin{figure*}[!htbp]
    \centering
    \includegraphics[width=1.0\textwidth]{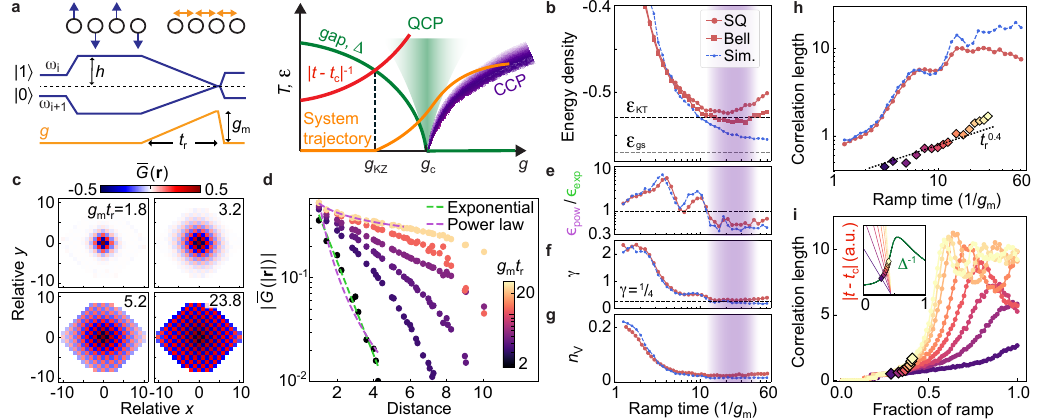}
    \caption{{\bf Quantum and classical critical coarsening in the XY-model. a,} Left: experimental schematic. The qubit frequencies (blue) are ramped from a staggered pattern to resonance while simultaneously turning on the coupling to $g_{\mathrm{m}} / (2 \pi)= 20$ MHz (yellow). Finally, qubits and couplers are ramped to idle levels, before measuring in the $X$- and $Y$-bases. Right: Phase diagram. The gap ($\Delta$) closes as $|g-g_{\text{c}}|^{0.67}$ (green). When the remaining time (red) to the critical point, $|t-t_{\text{c}}|$, exceeds $\Delta$, the dynamics become diabatic (dashed black), and the temperature ($T$) increases (orange). QCP and CCP: Quantum and classical critical phases. {\bf b,} The final energy density approaches the ground state value ($\varepsilon_{\mathrm{gs}}$; grey) and KT-transition value ($\varepsilon_{\mathrm{KT}}$; black) as $t_{\text{r}}$ is increased. Standard single qubit measurements (SQ; red circles) are limited by photon decay, which is corrected via Bell basis conversion (red squares). Blue: MPS simulation. Purple shading indicates where classical critical behavior is expected. {\bf c,} Average correlation, $\bar{G}(\bf{r})$ (found from averaging $(\langle X_iX_j+Y_iY_j\rangle-\langle X_i\rangle\langle X_j\rangle-\langle Y_i\rangle\langle Y_j\rangle)/2$ over all pairs $i,j$ separated by $\bf{r}$) measured at various $t_{\text{r}}$. {\bf d,} Decay of radially averaged correlations with Euclidean distance. Green and purple curves show examples of exponential and power law fits, respectively, performed up to maximum 6 sites to avoid finite-size effects at longer distances. {\bf e,} Ratio between rms errors from power law  and exponential fits ($\epsilon_{\mathrm{pow}}$ and $\epsilon_{\mathrm{exp}}$, respectively), showing notable increase in power-law character for $g_{\text{m}}t_{\text{r}}>15$. {\bf f,} Extracted power-law exponent, $\gamma$, decreases with $t_{\text{r}}$ and approaches expected value at KT-transition ($1/4$; black line). {\bf g,} Vortex density proxy, $n_{\mathrm{V}}$, decreases to minimum of $2\cdotp 10^{-2}$ in the low-energy regime. {\bf h,} Correlation length from exponential fits increases with $t_{\text{r}}$ and reaches 10 sites near $g_{\text{m}}t_{\text{r}}=25$. Both simulation results (blue) and experimental data (red) display substantially more superlinear growth than KZ predictions (black dashed). Correlation lengths extracted at expected freezing point (diamonds) agree better with KZ scaling (see {\bf i}). {\bf i,} Correlation length measured at intermediate times during the ramp. Diamonds represent expected freezing point, at which $\Delta=|t-t_{\text{c}}|$ (inset). The continued change in $\xi$ beyond the freezing point causes deviation from KZ predictions.}
    \label{fig:3}
\end{figure*}
A computationally challenging problem and useful benchmark for the quantum simulator is the thermalization dynamics of an initial $Z$-basis product state at half-filling, which has high temperature with respect to $H_{\text{s}}$ and hosts many quasiparticles (Fig.~\ref{fig:2}a). When subject to the (photon number conserving) time evolution operator $e^{-iH_{\text{s}} t/\hbar}$ where $\hbar$ is the reduced Planck constant (set to $1$ hereafter), interactions between quasiparticles are expected to drive the system into a chaotic state. To explore these dynamics, we perform a rapid (6 ns) ramp of the couplings $g_{ij} / 2 \pi$ from 0 to 10\,MHz. Quantum chaotic behavior is then diagnosed via $Z$-basis measurements at different times, yielding a set of probability distributions $p_{\rm meas}(x,t)$ where $\{ x \}$ represents the set of $D$ ``bitstrings'' with the same number of photons as the initial state. Figure~\ref{fig:2}b shows the distribution $\text{Pr}(p)$ of $p_{\rm meas}(x,t)$ for reduced system sizes up to $N_{\mathrm{q}} = 25$ at $t=100\text{ ns}\approx 6/g$. In each case, $\text{Pr}(p)$ exhibits a clear exponential decay known as the Porter-Thomas (PT) distribution, signalling thermalization to a quantum chaotic state~\cite{boixo_2018,shaw2024universal}. In contrast, past studies have found substantial deviations from the PT distribution in other models of analog dynamics~\cite{BenchmarkingSoonwonTheoryPRL2023,ShawNature2024}.

Characterizing the thermalization dynamics through the second moment of the bitstring distribution, also called the self-XEB~\cite{boixo_2018}, $D\sum_{x} p(x,t)_{\rm meas}^2-1$, we observe its fast convergence to the PT value of 1 within $t_{\mathrm{PT}} \approx 60$\,ns ($\sim 4/g$) for all system sizes (Fig.~\ref{fig:2}b inset, see SI for similar saturation rate of entanglement entropy). The observed fast scrambling dynamics are due to the simultaneously activated couplers and allow for reaching PT before decoherence causes substantial shifts towards a uniform $\text{Pr}(p)=D^{-1}$. Notably, the dynamics are approximately 4 times faster than in an equivalent digital circuit and thus less constrained by decoherence (SI). 

In order to also characterize the coherent errors from imperfect calibration of $H_{\text{s}}$, we consider the linear XEB fidelity, $F(t)=D\sum_x p_{\rm meas}(x,t)p_{\rm sim}(x,t)-1$, where $p_{\rm sim}$ are exactly simulated probabilities~\cite{boixo_2018}. The results, shown in Fig.~\ref{fig:2}c, exhibit exponential decay after times $\sim t_{\rm PT}$, where $F$ accurately describes the state fidelity (see SI for details). Fitting the decay, we obtain an error rate of $\epsilon = 0.10\pm0.02 \%$ per qubit per evolution time of $1/g$ (one cycle). Importantly, $\epsilon$ is nearly independent of system size up to the largest exactly-simulated system, $N_{\mathrm{q}} = 35$ (inset of Fig.~\ref{fig:2}c). This indicates the scalability of our calibration protocol and allows extrapolation to the full system size of $N_{\mathrm{q}} = 69$. Approximate matrix product state (MPS) simulations with bond dimension up to $\chi=1024$ were found to be ineffective beyond exactly simulatable system sizes, due to the fast entanglement growth and two-dimensional geometry of our system (SI). 

The combination of the observed fast dynamics and high fidelity enables quantum simulation of computationally complex states. A representative metric of this capability is the {\it mixed-state entanglement proxy}, $\mathcal{E}_{\text{P}}~=~S^{\text{Rényi-}\nicefrac{1}{2}}_{\text{ent}} + \log_2F$, which lower bounds the mixed-state entanglement by accounting for the effects of infidelity on the pure-state Rényi-$\nicefrac{1}{2}$ entropy~\cite{ShawNature2024}. Fig.~\ref{fig:2}d compares the estimated $\mathcal{E}_{\text{P}}$ of our work and other recent state-of-the-art experiments~\cite{AruteNature19,ZuchongzhiPRL21,ZuchongZhiSciBull2021,morvan2023phase,ShawNature2024}, where the proximity to the diagonal (ideal) line measures fidelity, indicating that our platform offers new possibilities for high-accuracy study of highly entangled states. In particular, we estimate that simulating quantum states to the level of our experimental fidelity requires over 1 million years on the Frontier supercomputer (SI).

\begin{figure*}[!htbp]
    \centering
    \includegraphics[width=1.0\textwidth]{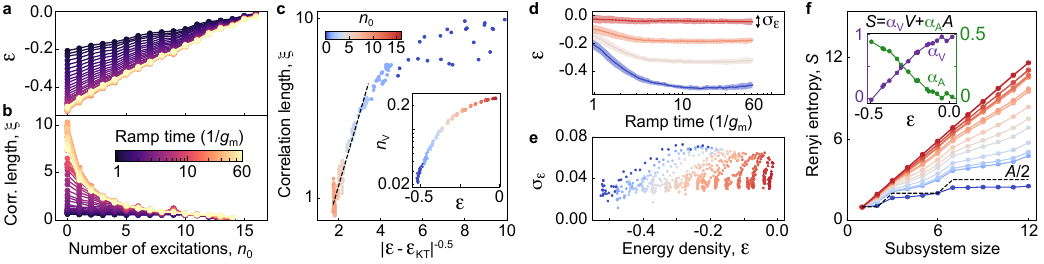}
    \caption{{\bf Tunable thermalized states via initial excitations. a,} Energy density and {\bf b,} correlation length as a function of number of initial excitations, $n_0$, averaged over 3 sets of randomized locations. {\bf c,} Main: Energy dependence of $\xi$, demonstrating collapse when data from sweeps of $t_{\text{r}}$ and $n_0$ are plotted together. $\log(\xi)$ is near-linear in $|\varepsilon-\varepsilon_{\mathrm{KT}}|^{-0.5}$ as theoretically expected. Inset: Vortex density vs energy density, showing similar collapse. {\bf d,} $t_{\text{r}}$-dependence of energy density and its fluctuations (width), for various $n_0$. {\bf e,} Energy fluctuations vs energy density, displaying absence of collapse since $\sigma_{\varepsilon}$ does not thermalize. {\bf f,} Main: Second Rényi entropy versus subsystem size for various $n_0$ at $t_{\text{r}}=200\text{\,ns}\approx 11.5/g$. Increasing $n_0$ causes transition from area- to volume-law behavior, also seen from the extracted ratio of their relative contributions (inset).}
    \label{fig:4}
    
\end{figure*}

Having explored the thermalization dynamics in the high-temperature regime, we next turn to the rest of the rich phase diagram in the XY-model (Eq.\,\eqref{eq:H}), which is expected to exhibit both a quantum phase transition in the ground state and a classical KT phase transition at finite temperature~\cite{KosterlitzThouless1973}. In order to prepare low-energy states of an antiferromagnetic XY magnet, we apply a staggered $z$-field of magnitude $h/(2\pi)=30$ MHz, and initialize the qubits in the $Z$-basis Neel state, maximizing the energy with respect to the first term in Eq.\,\eqref{eq:H}. We then ramp down the staggered field while simultaneously turning on {\it ferromagnetic} couplings of magnitude $g_{\mathrm{m}}/(2\pi)=20\,\text{MHz}$ over a duration $t_{\text{r}}$ (Fig.~\ref{fig:3}a). Under such a protocol~\cite{BrowaysXYNature2023}, the system evolution is equivalent to that of an antiferromagnetic XY model with staggered field, initialized in the ground state. This ramp crosses a quantum phase transition between a paramagnetic phase with unbroken $U(1)$ symmetry and the XY-ordered phase at $h_{\text{c}}/g_{\text{c}}\approx 1.8(6)$~(SI). The transition, analogous to the 2D Mott insulator-superfluid transition~\cite{FisherPRB1989}, is in the universality class of a 3D XY model, with the correlation length and dynamical critical exponents $\nu\approx 0.67$ and $z=1$, respectively. The classical KT transition temperature in the XY-ordered phase vanishes at the quantum critical point at $g_{\text{c}}$ as $T_{\rm KT}\sim (g-g_{\text{c}})^{\nu}$. Following the ramp, we rapidly return back to the idle frequencies within 3 ns and perform measurements of correlation functions. 

Figure~\ref{fig:3}b shows the ramp time dependence of the average energy density, $\varepsilon=n_{\mathrm{B}}^{-1}\sum_{\langle i,j\rangle}\langle X_iX_{j}+Y_iY_j\rangle/2$ averaged over $n_{\mathrm{B}}=110$ bonds ($N_{\mathrm{q}}=65$) and corrected for readout errors (SI). (Note that $\varepsilon$ is made dimensionless through normalization by $g_{\mathrm{m}}$). As $t_{\text{r}}$ increases and the dynamics become more adiabatic, we observe a decrease in energy density toward the theoretically predicted ground state value of $\varepsilon_{\text{gs}}=-0.56$, as well as the predicted KT-transition energy density, $\varepsilon_{\text{KT}}=-0.53\,\pm\,0.01$ (grey and black dashed lines, respectively). Importantly, as we shall demonstrate below, the final states are thermalized to a strong extent, so the measured energy can be used to evaluate the final effective temperature. To correct for photon decay errors, we apply digital entangling gates at the end of the circuit to convert each pair of qubits to the Bell basis (SI). This allows for postselecting with respect to photon number conservation, which yields an improved value of $\varepsilon=-0.53\pm 0.01$, approximately equal to the KT-transition point (red squares). The remaining discrepancy from $\varepsilon_{\text{gs}}$ likely arises from dephasing effects, which are not corrected by this technique.

Since the energy itself does not reveal the effects of thermalization, we next turn to correlations at longer distances and consider the average correlation, $\bar{G}(\bf{r})$, between pairs of qubits separated by $\bf{r}$, shown in Fig.~\ref{fig:3}c. We observe antiferromagnetic ordering, with the range and magnitude of correlations increasing dramatically with ramp time, as expected for states with decreasing energy. We next compute the radial average, $\bar{G}(|\bf{r}|)$, and fit the resulting decay profiles with exponential fits to extract the correlation length, $\xi$, as well as with power-law fits to evaluate the type of distance-scaling (Fig.~\ref{fig:3}d). At short ramp times, the correlations are found to decay exponentially, as theoretically expected for states above the KT-transition, where freely proliferating vortices preclude long-range order. At longer ramp times, on the other hand, the decay behavior is better described by power-law fits, as shown in Fig.~\ref{fig:3}e; specifically, we observe a marked decrease in the ratio between the root-mean-square errors of power-law and exponential fits to well below 1 near $g_{\text{m}}t_{\text{r}}=25$, where the energy is also close to its minimum value. This behavior is consistent with that expected in the classical critical regime, where free vortices become entropically unfavorable and are replaced by bound vortex-antivortex pairs, leading to algebraically decaying correlations. (We note that finite-size scaling analysis of the KT transition is challenging, due to characteristic rapid growth of the correlation length, and we do not attempt it here.) In the region with good power-law agreement, we extract a power-law exponent of $\gamma=-0.29$ (Fig.~\ref{fig:3}f), close to the theoretically expected universal value of $-\frac{1}{4}$ at the KT-transition~\cite{NelsonKosterlitzPRL1977}.

In order to further substantiate our interpretation, we also measure 4-qubit correlators to construct the Swendsen proxy for the vortex density~\cite{swendsen1982first}, given by $n_{\mathrm{V}}~=~\frac{1}{4N_{\mathrm{P}}}\sum_{\mathrm{i=1}}^{N_{\mathrm{P}}}(1-X_{i1}X_{i3}-Y_{i2}Y_{i4})(1-Y_{i1}Y_{i3}-X_{i2}X_{i4})$ for plaquettes $i=1,..,N_{\mathrm{P}}$ with vertices $\{i1,i2,i3,i4\}$. Indeed, we observe a rapid decrease in ${n}_{\mathrm{V}}$ as $t_{\text{r}}$ is increased (Fig.~\ref{fig:3}g), reaching a minimum value of $2\cdotp 10^{-2}$ in the low-energy regime.

\begin{figure*}[!htbp]
    \centering
    \includegraphics[width=1\textwidth]{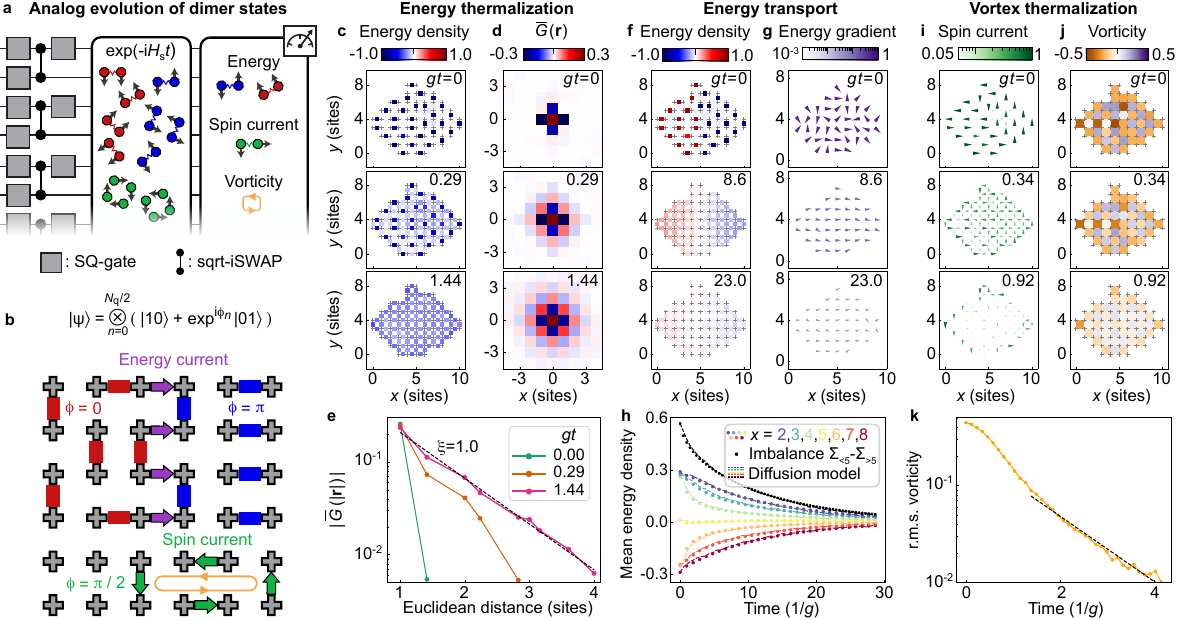}
    \caption{{\bf Transport and thermalization dynamics with entangled initial states. a,} Dimer states are prepared using digital gates, and their thermalization and transport dynamics are realized with analog evolution, before finally measuring energy, spin current and vorticity. {\bf b,} We prepare dimer states with spatially tunable phase, $\phi$. Energy gradients between $\phi=0$ ($\varepsilon>0$) and $\phi=\pi$ ($\varepsilon<0$) drive energy current, while $\phi=\pi/2$ gives non-zero spin-current and vorticity. {\bf c,} Time evolution of energy density and {\bf d,} correlations after dimer preparation demonstrate rapid thermalization. {\bf e,} Correlations become increasingly long-ranged as the system thermalizes. Dashed line: exponential fit. {\bf f,} Energy density and {\bf g,} energy gradients after dimer preparation with $\phi=0$ and $\pi$ in the left and right halves of the system, respectively, displaying energy transport on much longer timescales. Color and length scales of arrows in {\bf g} and {\bf i} are logarithmic. {\bf h,} Time dependence of average energy density along various vertical cuts (colored circles) and energy imbalance across $x=5$ (black circles), exhibiting very good agreement with diffusion model (dashed lines). {\bf i,} Spin current and {\bf j,} vorticity for $\phi=\pi/2$, exhibiting rapid thermalization. {\bf k,} The r.m.s. vorticity shows initial slow dynamics followed by near-exponential decay with rate $\Gamma=49\,\text{MHz}=0.85\,g$ (fit shown by dashed line).}
    \label{fig:5}
\end{figure*}

Having studied the classical critical behavior, we next explore its effects on the scaling of the correlation length with the duration $t_{\text{r}}$ over which we sweep through the quantum critical point (Fig.~\ref{fig:3}h). The correlation length rises to a maximum of $\xi\sim 10$ at $g_{\mathrm{m}}t_{\text{r}}=25$, which is equal to the longest dimension of our system. At long ramp times, we observe a slight decrease in $\xi$, attributed to qubit decoherence, as well as periodic oscillations. The latter are also observed in MPS simulations and likely due to finite-size effects. Focusing on shorter ramp times where these additional effects are absent, we observe strong deviation from the power-law scaling with exponent $\nu/(1+\nu z)=0.4$ predicted by KZ theory ($\nu=0.67,z=1$). Specifically, $\xi$ grows substantially more superlinearly, and clear discrepancies from power-law scaling are observed in both experiment and simulation. We attribute the observed breakdown of KZ scaling to both quantum and classical critical coarsening~\cite{SamajdarArxiv2024}. 

In order to demonstrate this more explicitly, we proceed to measure the correlation length along the Hamiltonian ramp (Fig.~\ref{fig:3}i). The KZ prediction assumes that the dynamics freeze when the inverse gap, $\Delta^{-1}\propto |g-g_{\text{c}}|^{-\nu z}$, exceeds the remaining time of the ramp (marked as diamonds). In contrast, we find that $\xi$ continues to increase, suggesting that the system is instead able to further thermalize, thus giving rise to a different correlation length when measured at the end of the ramp. This effect is amplified by the presence of the classical critical region, in which the correlation length becomes much longer during thermalization than what it was at the more naive freezing point. To illuminate this discrepancy further, we plot the experimentally measured correlation lengths at the theoretically predicted freezing points in Fig.~\ref{fig:3}h and find better agreement with the KZ prediction. 

Thus far, we have tuned the energy scale of our system via the ramp rate of the Hamiltonian. To further study thermalization, as well as the scaling relations near the KT-transition, we prepare a variable number of excitations, $n_0$ (pairs of spin flips in randomized locations) in the initial state~\cite{schuckert2023observation}. While we find that the final average energy density depends linearly on $n_0$ (Fig.~\ref{fig:4}a), the behavior of the correlation length is more intricate (Fig.~\ref{fig:4}b) and is best understood by plotting $\xi$ versus energy density for all $g_{\mathrm{m}}t_{\text{r}}>5$ and $n_0$ (Fig.~\ref{fig:4}c). Notably, the points exhibit a collapse (also observed for $n_{\text{V}}$, see inset), suggesting that the final states are well thermalized, such that the energy density of the final state determines $\xi$ and $n_{\text{V}}$, as expected from the eigenstate thermalization hypothesis (ETH)~\cite{DeutschETH,SrednickiETH}. Barring $\xi$ near the system size, we find that $\log\xi$ is nearly linear in $|\varepsilon-\varepsilon_{\mathrm{KT}}|^{-0.5}$, as is predicted near the KT-transition. This is incompatible with naive KZ scaling, and thus further corroborates that classical-critical coarsening can preclude the KZ mechanism.

While thermalization causes states created with different $n_0$ and $t_{\text{r}}$ to have the same observables (e.\,g. $n_{\mathrm{V}}$ and $\xi$) if their final energy is identical, the states themselves are not necessarily the same. This can be seen by studying observables such as the energy fluctuations, $\sigma_{\varepsilon}=(n_{\mathrm{B}}g_{\mathrm{m}})^{-1}\sqrt{\langle H_{XY}^2\rangle-\langle H_{XY}\rangle^2}$ with $H_{XY}=\sum_{\langle i,\,j \rangle}(X_iX_j+Y_iY_j)/2$, which trivially commute with the Hamiltonian and are thus not thermalized under ETH. We next reconstruct $\sigma_{\varepsilon}$ from 2- and 4-qubit correlators (SI) and find that it decreases from $\sim 0.07$ to $\sim 0.02$ as we approach the ground state for $n_0 = 0$, while its dependence on $n_0$ is much weaker (Fig.~\ref{fig:4}d). The very low value of $\sigma_{\varepsilon}$ compared to the tunable energy range indicates our ability to probe  specific parts of the spectrum. Notably, when the full dataset across $t_{\text{r}}$ and $n_0$ is plotted against energy density, the points do not collapse (Fig.~\ref{fig:4}e). This reveals the difference in states accessed by the two tuning techniques, which was previously concealed by the thermalization of other observables.

To further characterize the degree of thermalization, we leverage the fast data acquisition rate of our platform to measure the entanglement entropy for subsystem sizes up to 12 qubits, using randomized measurements\cite{brydges2019probing}. At $n_0=0$, we find area-law behavior (Fig.~\ref{fig:4}f), which, up to a subleading logarithmic  contribution, is consistent with predictions for low-energy states in the XY model~\cite{metlitski2015entanglement}. However, tuning to higher final energies via larger $n_0$, we find a continuous crossover to volume-law behavior (area- and volume-law components in inset), as is expected from ETH for thermalized states at finite energy density~\cite{AbaninRMP2019,karamlou_2023}.

We have so far observed signatures of thermalization in the final state of the dynamics, but the thermalization dynamics themselves are still left unexplored. While we have shown that $t_{\text{r}}$ and $n_0$ are effective for realizing and studying states with a desired energy and energy fluctuations, they are limited when it comes to studying spatiotemporal dynamics; in order to study a state with substantial correlations ($\langle XX\rangle>0.1$), a ramp time of more than $\sim 1/g$ is required, at which point the system is typically already near equilibrium. Moreover, while these knobs allow for tuning energy density and antiferromagnetic correlations, quantities like vorticity are out of reach. 

Next, we therefore expand the capabilities of our platform by combining the analog evolution with entangled state preparation via high-fidelity (digital) two qubit-gates (Fig.~\ref{fig:5}a,b). Following the preparation of the dimer state, $(\ket{01}-\ket{10})^{\otimes N_{\mathrm{q}}/2}$, we rapidly turn on $H_s$ with $g/(2\pi)=10$ MHz and observe very fast thermalization of the energy density on a timescale of just $\sim 1.5/g$ (Fig.~\ref{fig:5}c). As the system thermalizes, the range of correlations increases rapidly (Fig.~\ref{fig:5}d), converging to a correlation length of $\sim 1.0$ (Fig.~\ref{fig:5}e). As is expected from ETH, this is in good agreement with $\xi\sim 1.1$ observed for the same energy  density ($-0.23g$) when tuning $t_{\text{r}}$ and $n_0$.

Next we leverage the tunability of the phases of the initial dimer states to enable slower dynamics and study of transport (Fig.~\ref{fig:5}b). Specifically, we now prepare the dimers in one half of the device in the higher-energy dimer state, $\ket{01}+\ket{10}$ (Fig.~\ref{fig:5}f). Now the dynamics are found to be substantially slower, with clear spatial non-uniformity remaining even after 23 cycles. We also plot the energy density gradient in Fig.~\ref{fig:5}g, which quickly establishes a uniform field in the $+x$-direction. Fig.~\ref{fig:5}h shows the time dependence of the average energy density at various vertical cuts (colored circles), as well as the total energy transfer across $x=5$ (black circles), which both exhibit excellent agreement with a diffusion model (dashed lines). The energy transport is indeed expected to be diffusive in this regime, due to the relatively high energy of the dimer state. The data allows for extracting a diffusion constant of $D$ = $29.6\,\text{MHz} = 0.52\,g$.

The use of initial entangled states in our hybrid analog-digital platform enables not only tailoring the initial energy landscape, but also other observables such as vorticity and spin current. We achieve this by further tuning the initial dimer phases to $\pi/2$ (Fig.~\ref{fig:5}b). This gives rise to local spin currents, $\langle X_iY_{i+1}-Y_iX_{i+1}\rangle/2\neq0$, and a sea of vortices and anti-vortices, quantified by the vorticity, $V_i=\frac{1}{4}(X_{i1}Y_{i2}-Y_{i2}X_{i3}+X_{i3}Y_{i4}-Y_{i4}X_{i1})$ for each plaquette $i$ with vertices $\{i1,i2,i3,i4\}$. The temporal evolution of the spin current and vorticity is presented in Figs.~\ref{fig:5}i and j, respectively, showing thermalization on a fast timescale similar to that in Fig.~\ref{fig:5}c. Specifically, after an initial super-exponential decay, the root-mean-square vorticity decays near-exponentially with a rate of $\Gamma= 49\,\text{MHz} = 0.85\,g$ (Fig.~\ref{fig:5}k).

Our results demonstrate a high-fidelity quantum simulator with the capability of emulating beyond-classical chaotic dynamics, a wide range of characterization probes, and versatile analog-digital control. Leveraging these features has enabled new insights about the rich interplay of quantum and classical critical behavior in the 2D XY-model, including the KT transition, thermalization dynamics, and their combined effects on the KZ scaling relations. Looking ahead, this platform is expected to offer an invaluable playground for studies of classically intractable many-body quantum physics, including e.g. dynamical response functions and magnetic frustration. 

\textit{Note} -- During the preparation of this manuscript, the authors became aware of a related work studying coarsening near an Ising quantum phase transition with Rydberg atoms~\cite{manovitz2024quantum}.

\vspace{1cm}


\subsection*{Acknowledgments} The authors thank Soonwon Choi for helpful discussion. A.S. acknowledges support from the U.S. Department of Energy, Office of Science, National Quantum Information Science Research Centers, Quantum Systems Accelerator. A.E. acknowledges funding by the German National Academy of Sciences Leopoldina under the grant number LPDS 2021-02 and by the Walter Burke Institute for Theoretical Physics at Caltech. Work in Grenoble is funded by the French National Research Agency via the JCJC project QRand (ANR-20-CE47-0005), Laboratoire d’excellence LANEF (ANR-10-LABX-51-01), from the Grenoble Nanoscience Foundation.

\subsection*{Author contributions}
T.I. Andersen, D.A. Abanin and X. Mi conceived the project and designed the experiments. T.I. Andersen, A. Karamlou and J. Berndtsson performed the experiments and data analysis. T.I. Andersen, J.A. Gross, X. Mi and D.A. Abanin developed the calibration procedures, with assistance from N. Astrakhantsev, Y. Zhang, E. Forati, B. Kobrin, A. Di Paolo, A.R. Klots, I. Drozdov, A. Petukhov and L.B. Ioffe. J. Motruk, A. Szasz, D. Rossi and D.A. Abanin performed MPS simulations and theoretical work with A.M. Lauchli. N. Astrakhantsev performed XEB analysis, classical complexity estimates and exact state vector simulations with assistance from T. Westerhout, V.D. Kurilovich and A. Szasz. A. Elben, A. Rath, V. Vitale and B. Vermersch performed theoretical work on randomized measurements. T.I. Andersen, N. Astrakhantsev, A.M. Lauchli, D.A. Abanin and X. Mi wrote the manuscript. D.A. Abanin and X. Mi led and coordinated the project. Infrastructure support was provided by Google Quantum AI. All authors contributed to revising the manuscript and the Supplementary Materials.

\clearpage
\onecolumngrid
\vspace{1em}

\twocolumngrid
\pagebreak
\pagebreak
\clearpage

\newcommand\SupplementaryMaterials{
  \xdef\presupfigures{\arabic{figure}}
  \xdef\presupsections{\arabic{section}}
  \renewcommand\thefigure{S\fpeval{\arabic{figure}-\presupfigures}}
  \renewcommand\thesection{S\!}
 
}
\newpage
\pagebreak
\clearpage

\SupplementaryMaterials

\section*{Supplementary materials}
\subsection{Device details}
\begin{figure}[!htbp]
    \centering
    \includegraphics[width=\columnwidth]{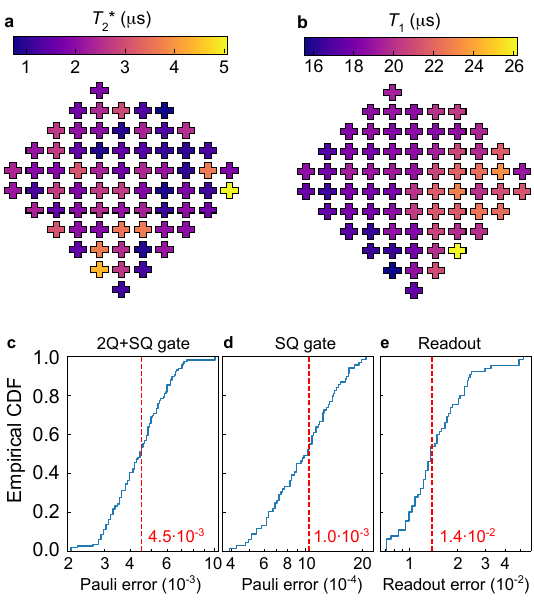}
    \caption{{\bf Device characterization.} {\bf a,b,} Ramsey dephasing ($T_2^*$; {\bf a}) and photon relaxation ($T_1$; {\bf b}) times across the qubit grid. {\bf c,d,} Histogram of Pauli error for combined $\sqrt{\text{iSWAP}}$ and single qubit gates ({\bf c}) and only single qubit gates ({\bf d}). Red dashed lines indicate the median values. (CDF: cumulative distribution function). {\bf e,} Histogram of readout errors.}
    \label{fig:SIA}
\end{figure}

The experiments are performed on a superconducting quantum processor with frequency-tunable transmon qubits and couplers, with a similar design to that in Ref.~\cite{AruteNature19}. Fig.~\ref{fig:SIA}a,b show the measured Ramsey dephasing ($T_2^*$) and photon relaxation ($T_1$) times at the  interaction frequency of $5.93$\,GHz used in our experiments, with median values of $2.0\,\mu \text{s}$ and $18.8\,\mu\text{s}$, respectively. Characterizing our digital gate performance, we find a median Pauli error of $4.5\times 10^{-3}$ for combined $\sqrt{\text{iSWAP}}$ and single-qubit gates (Fig.~\ref{fig:SIA}c), and $1.0\times 10^{-3}$ for single qubit gates alone (Fig.~\ref{fig:SIA}d). Finally, Fig.~\ref{fig:SIA}e displays our readout errors, with a median of $1.4\times 10^{-2}$.

\subsection{Analog calibration}
In this section, we describe our new, scalable analog calibration framework that enables $\sim 0.1\%$ cycle error per qubit. In order to achieve a scalable scheme, we perform pairwise calibration measurements --- specifically single-photon and swap spectroscopy --- which allows for accurately setting the effective coupling $\tilde{g}$ and dressed qubit frequencies $\tilde{\omega}_{qi}$ in each qubit pair. A key challenge in analog calibration that contrasts with its digital counterpart is that these dressed quantities in the pairwise scenario change drastically when all couplers are turned on in the fully-coupled global case. Therefore, we perform extensive modeling of the device physics to accurately convert them to the bare qubit and coupler frequencies, $\{\omega_{qi}\}$,$\{\omega_{cj}\}$, which --- crucially --- do not change from the local calibration measurements to the full-scale experiments. 

\subsubsection{Model device Hamiltonian}
\begin{figure}[!htbp]
    \centering
    \includegraphics[width=\columnwidth]{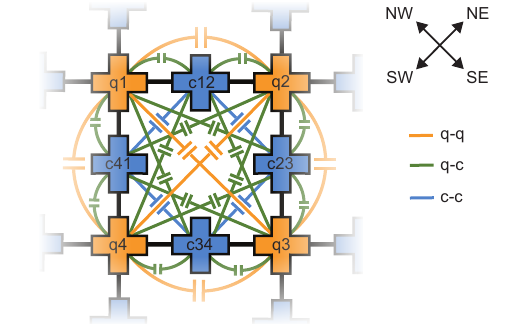}
    \caption{{\bf Schematic of underlying coupling pathways in the device.} In addition to capacitive coupling between neighboring qubits (orange) and couplers (blue), there are also diagonal next-nearest-neighbor couplings. Asymmetry in the underlying structure of the qubits causes a difference in the couplings along the NW-SE and NE-SW diagonals.}
    \label{fig:SI1}
\end{figure}

We model both the qubits and couplers in our tunable coupler architecture as Kerr oscillators, with 4 or 5 levels in each transmon, depending on the number of photons involved in the Hamiltonian term of interest. In order to ensure high accuracy, we account for not only coupling terms between neighboring qubits and couplers, but also diagonal pathways, including between couplers:
\begin{align}
 H_d=\nonumber&\overbrace{\sum_{qi} \omega_{qi} \hat{n}_{qi}-\eta_{qi}\hat{n}_{qi}(\hat{n}_{qi}-1)/2}^{\text{Single qubit}}+\\
 \nonumber&\overbrace{\sum_{cj} \omega_{cj} \hat{n}_{cj} -\eta_{cj}\hat{n}_{cj}(\hat{n}_{cj}-1)/2}^{\text{Single coupler}}+\\
 \nonumber&\overbrace{\sum_{qi,qj}\frac{1}{2}\tilde{k}_{qi,qj}\sqrt{\omega_{qi}\omega_{qj}}\hat{Q}_{qi}\hat{Q}_{qj}}^{\text{Qubit-qubit coupling}}+\\
 \nonumber&\overbrace{\sum_{qi,cj}\frac{1}{2}\tilde{k}_{qi,cj}\sqrt{\omega_{qi}\omega_{cj}}\hat{Q}_{qi}\hat{Q}_{cj}}^{\text{Qubit-coupler coupling}}+\\
&\overbrace{\sum_{ci,cj}\frac{1}{2}\tilde{k}_{ci,cj}\sqrt{\omega_{ci}\omega_{cj}}\hat{Q}_{ci}\hat{Q}_{cj}}^{\text{Coupler-coupler coupling}},
\end{align}
where $\hat{Q}=a^{\dagger}+a$ and the $\tilde{k}$ are the effective coupling efficiencies between transmons, including both direct and indirect capacitive contributions (note that the indirect contributions should not be confused with contributions due to virtual exchange interactions, which are included indirectly when we project out the couplers later on). The coupling efficiencies for the various terms can be summarized as follows:
\newline
$\bf{k_{qq}}$: We include three types of qubit-qubit coupling, distinguished by the relative positioning of the qubits. Notably, the geometry of the transmons break the symmetry between the northwest-southeast (NW-SE) and northeast-southwest (NE-SW) directions. To discuss the three types of coupling, we consider the 4 qubits on a plaquette shown in Fig.~\ref{fig:SI1}: \newline
1) Nearest-neighbors qubits, $q1$ and $q2$ separated by a coupler $c12$: $\tilde{k}_{q1,q2}=k_{q1,q2}+k_{q1,c}k_{q2,c}$. \newline
2) Diagonally separated qubits in the NW-SE direction, $q1$ and $q3$: $\tilde{k}_{q1,q3}=k_{q1,q3}+2(k_{q1,q2}k_{q2,q3}+k_{q1,q4}k_{q4,q3})$. \newline
3) Diagonally separated qubits in the NE-SW direction, $q2$ and $q4$: $\tilde{k}_{q1,q3}=k_{q1,q3}$.

$\bf{k_{qc}}$: We also include three types of qubit-coupler coupling: \newline
1) Nearest-neighbors: $\tilde{k}_{q1,c1}=k_{q1,c1}$. \newline
2) Diagonally separated qubit and coupler in the NW-SE direction, $q1$ and $c23$: $\tilde{k}_{q1,c23}=k_{q1,c23}+2k_{q1,q2}k_{q2,c23}$. \newline
3) Diagonally separated qubit and coupler in the NE-SW direction, $q4$ and $c12$: $\tilde{k}_{q4,c12}=k_{q4,c12}$.

$\bf{k_{cc}}$: Finally, we consider two types of coupler-coupler coupling:\newline
1) Diagonally separated couplers in the NW-SE direction $c12$ and $c23$:
$\tilde{k}_{q1,c23}=k_{c12,c23}+2k_{c12,q2}k_{q2,c23}$.\newline
2) Diagonally separated qubit and coupler in the NE-SW direction, $c12$ and $c14$: $\tilde{k}_{c12,c14}=k_{c12,c14}$.

\subsubsection{Calibration experiments}

\begin{figure}[!htbp]
    \centering
    \includegraphics[width=\columnwidth]{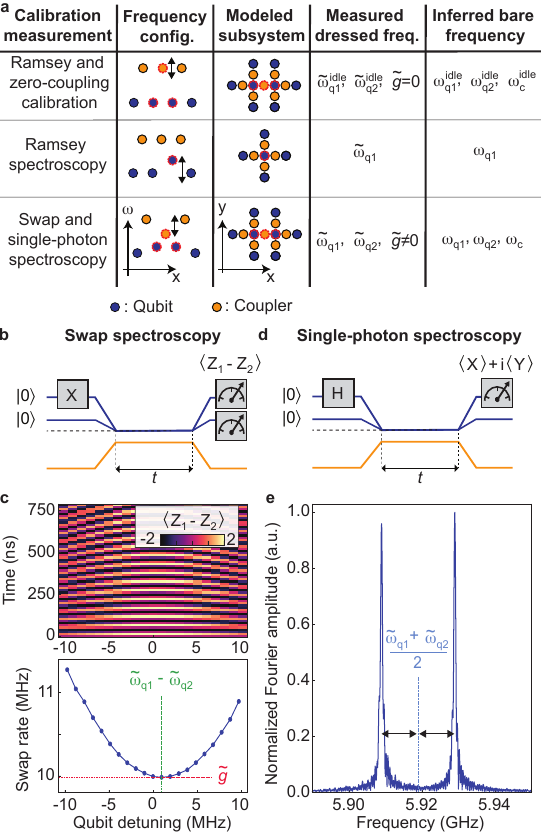}
    \caption{{\bf Analog calibration procedure. a,} Overview of calibration steps. We perform three main steps, which together allow for determining the bare frequencies of the qubits and couplers in the idle configuration in which $\tilde{g}=0$ (top row), as well as in the interaction configuration (bottom two rows). For each step, we model a subsystem (third column) to convert the measured dressed frequencies (fourth column) to bare frequencies (fifth column). {\bf b,} Circuit schematic of swap spectroscopy. {\bf c,} Top: Measured population difference, $\langle Z_1-Z_2\rangle$, as a function of qubit detuning and time. Bottom: Extracted swap rate from Fourier transform vs qubit detuning. The position of the minimum allows for determining $\tilde{g}$ and the difference of the dressed qubit frequencies, $\tilde{\omega}_{q1}-\tilde{\omega}_{q2}$. {\bf d,} Circuit schematic of single-photon spectroscopy. {\bf e,} Fourier transform of the measured $\langle X\rangle+i\langle Y\rangle$. The average of the peak positions is equal to the average of the dressed qubit frequencies $(\tilde{\omega}_{q1}+\tilde{\omega}_{q2})/2$.}
    \label{fig:SI2}
\end{figure}

\begin{figure*}[!tbp]
    \centering
    \includegraphics[width=\textwidth]{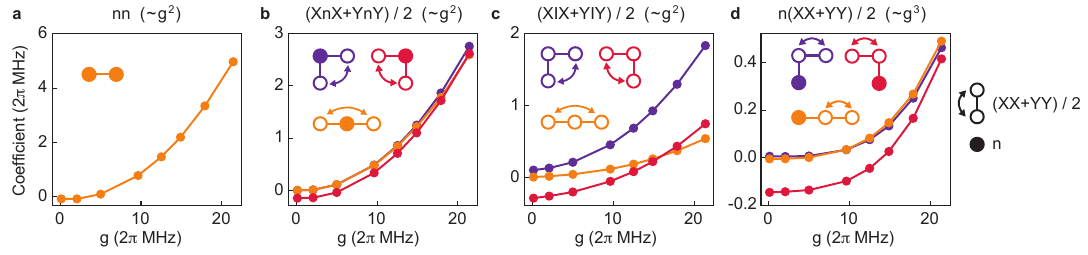}
    \caption{{\bf Higher order terms in the analog spin Hamiltonian.} Average coupling coefficient vs nearest-neighbor hopping $g$ for {\bf a,} $n_in_{i+1}$, {\bf b,} $(X_in_{i+1}X_{i+2}+Y_in_{i+1}Y_{i+2})/2$, {\bf c,} $(X_iX_{i+2}+Y_iY_{i+2})/2$, and {\bf d,} $n_i(X_{i+1}X_{i+2}+Y_{i+1}Y_{i+2})/2$, where qubits $i$, $i+1$, and $i+2$ are placed along a connected line. The first three terms scale as $g^2/\eta$, while the fourth scales as $g^3\eta^2$, where $\eta$ is the anharmonicity. At $g=2\pi\times10\text{ MHz}$, all higher-order terms are smaller than $1\times2\pi$ MHz. In the three latter terms, there is asymmetry between the three possible configurations displayed in the insets (see text for details). Note that $n_i(X_{i+1}X_{i+2}+Y_{i+1}Y_{i+2})/2$ does not differ on average from $(X_{i}X_{i+1}+Y_{i}Y_{i+1})n_{i+2}/2$ in {\bf d}.}
    \label{fig:SI3}
\end{figure*}
In order to calibrate the bare qubit and coupler frequencies for a given set of applied biases, we perform various types of calibration measurements (Fig.~\ref{fig:SI2}a):\newline
\textbf{Ramsey spectroscopy}:
In this measurement, we perform standard Ramsey spectroscopy for a range of applied qubit bias values, while keeping the couplers turned off and the neighboring qubits detuned, in order to prevent swapping. \newline

\textbf{Swap spectroscopy}:
This measurement is performed on a pairwise level, where neighboring couplers (except the one connecting the pair) are turned off. The two qubits are prepared in the $\ket{10}$-state and we measure the swap rate as a function of detuning between the two qubits (Fig.~\ref{fig:SI2}b). The minimum swap rate tells us the effective coupling between the two qubits, $\tilde{g}$, and the detuning at which this occurs equals the difference between the dressed frequencies of the qubits, $\tilde{\omega}_{q1}-\tilde{\omega}_{q2}$ (Fig.~\ref{fig:SI2}c). Using an iterative scheme, we calibrate the coupler bias required to achieve the target effective coupling.\newline

\textbf{Single-photon spectroscopy}:
While the swap spectroscopy provides us with the difference of the dressed frequencies, we also need to find their sum to determine the individual values, $\tilde{\omega}_{q1}$ and $\tilde{\omega}_{q2}$. We achieve this by preparing the qubits in $(\ket{1}+\ket{0})\ket{0}/\sqrt{2}$ and measuring $\langle X+iY\rangle$ as a function of evolution time (Fig.~\ref{fig:SI2}d). The Fourier transform of the signal then reveals the eigenfrequencies of the two-qubit system, the average of which is equal to $(\tilde{\omega}_{q1}+\tilde{\omega}_{q2})/2$ (Fig.~\ref{fig:SI2}e). 

Next, using separately calibrated coupling efficiencies, we model all the calibration experiments above with the device Hamiltonian described earlier, in order to find the bare qubit and coupler frequencies that give the dressed quantities observed in the calibration experiment. Importantly, we model not only the two qubits and the coupler involved in pairwise experiments (single qubit involved in Ramsey), but also the neighboring ``padding'' qubits and couplers in order to account for their effects. Therefore, we start by determining the bare \textit{idle} frequencies, $\{\omega^{\text{idle}}\}$, since these must be known to represent the ``padding'' in the interaction configuration. 

\subsubsection{Projection onto computational subspace}
Considering the fact that our model device Hamiltonian involves both qubits and couplers with up to 5 levels in each, it is computationally intractable to use it for time evolution even at small photon numbers. Moreover, in this form, it is very difficult to map its behavior onto physically relevant systems. We therefore perform a projection technique to convert the device Hamiltonian into a spin-Hamiltonian, $H_s$, that acts on the computational subspace. To find spin-Hamiltonian terms involving $n$ photons in a system of $N_{\mathrm{q}}$ qubits, we write $H^{(n)}=\sum_{i,j}\ket{i}\bra{i}H_d\ket{j}\bra{j}$, where $\{\ket{i}\}$ are our $N_n={N_{\mathrm{q}} \choose n}$ new dressed $n$-photon basis states. 

Let us now motivate our choice of dressed basis states, by considering a few different options. One option could have been to simply use the bare qubit states, $\{\ket{i}_{\mathrm{bare}}\}$; however, this would cause the spin-Hamiltonian to have different eigen-energies from the low-energy spectrum of $H_d$. A second option would be to instead use the $N_n$ lowest-energy $n$-photon eigenstates of $H_d$, $\{\ket{i}_{\mathrm{eigen}}\}$. In this case, the spin Hamiltonian is guaranteed to have the same $N_n$ lowest $n$-photon eigen-energies as $H_d$. However, these basis states are highly delocalized and poorly represent our qubits. Hence, to get the best of both worlds, we turn to a third option, where we project the bare qubit states onto the low-energy eigenspace spanned by $\{\ket{i}_{\mathrm{eigen}}\}$. These projections are not orthonormal, so we perform singular value decomposition and set the singular values to 1 in order to arrive at our new dressed basis states. It can be shown that this is the most localized set of states that still preserve the low-energy eigenvalues~\cite{bravyi2011schrieffer}. These new basis states are slightly delocalized on the nearest couplers and qubits, and also have a weak overlap with states that have $n+2$ and $n-2$ photons due to terms beyond the rotating-wave approximation. We note that our typical coupler ramp times of $>5$ ns are sufficient to ensure adiabatic conversion between the bare qubit states (in which we perform state preparation and measurement) and the dressed basis states that are relevant under analog evolution.

The spin-Hamiltonian $H^{(n)}$ found from the technique above in principle includes all terms involving $\leq n$ photons, including very long-range interactions; however, they drop off rapidly with the photon-photon separation $d$ (typically as $(g/\eta)^d\sim0.1^d$). Moreover, we also find that the terms decay with the number of involved photons in a similar way. Hence, in order to achieve the low error demonstrated in our manuscript, it is sufficient to include only terms involving up to 2 photons, and where all the involved qubits are a maximum Manhattan distance of 2 sites apart, resulting in:
\begin{align}
\label{eq:higher_order}
 \nonumber&H=\sum_i \omega_i n_i+\sum_{\langle i,\,j\rangle} g_{ij}(X_iX_j+Y_iY_j)/2+\sum_{\langle i,\,j\rangle}g^{nn}_{ij}n_in_j\\
 \nonumber&+\sum_{\langle i,\,j,\,k\rangle}(g^{XnX}_{ijk}n_j+g^{XIX}_{ijk})(X_iX_k+Y_iY_k)/2\\ 
&+\sum_{\langle i,\,j,\,k\rangle}(g^{nXX}_{ijk}n_i(X_jX_k+Y_jY_k)/2,
\end{align}
where $g^{nn}_{ij}$, $g^{XnX}_{ijk}$ and $g^{XIX}_{ijk}$ scale as $g^2/\eta$, while $g^{nXX}_{ijk}$ scales as $g^3/\eta^2$, and qubits $i,j,k$ are connected (see Fig.~\ref{fig:SI3}).

Our technique requires finding the $N_n$ lowest-energy $n$-photon eigenstates of $H_d$, which has a high computational cost for large $N_{\mathrm{q}}$. Fortunately, for a given Hamiltonian term involving a certain set of qubits, the effect of other transmons decays quickly with distance, and we only need to include the nearest neighboring qubits and couplers to achieve accuracies on the tens of kHz scale. To find the spin-Hamiltonian terms, we therefore scan through various subsystems and perform the procedure outlined above for each of them. 

\subsection{Phase calibration for hybrid analog-digital experiments}

\begin{figure*}[!htbp]
    \centering
    \includegraphics[width=1\textwidth]{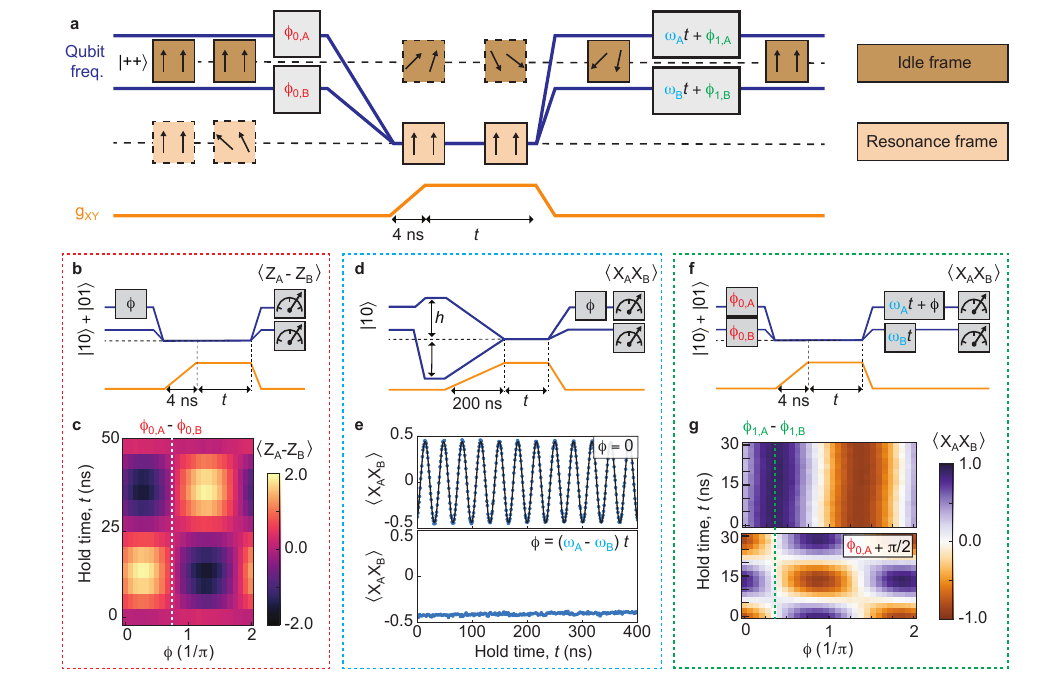}
    \caption{{\bf Phase calibration for hybrid analog-digital experiments. a,} Schematic of phase accumulation and correction throughout hybrid analog-digital circuit. While we typically prepare initial dimer states, we here consider an initial state $\ket{++}$ for the purpose of simplified explanation. Blue and yellow lines show qubit frequency trajectories and coupling profile, respectively, while brown (beige) boxes show the relative alignment of the two spins in the idle (resonance) frame. We apply corrective phases $\{\phi_{0,i}\}$ before the analog circuit to ensure the correct dimer phase in the resonance frame when the analog Hamiltonian is turned on. Additional phases $\{\omega_{0,i} t+\phi_{1,i}\}$ are applied after the analog evolution in order to measure the same phase in the idle frame as was in the resonance frame. {\bf b,} $\{\phi_{0,i}\}$ are calibrated by preparing triplet states, sweeping the phase difference within each qubit pair, and measuring the population difference after a variable time $t$. {\bf c,} Population difference after time $t$ for an applied phase difference $\phi$. Since only the dimer phases $0$ and $\pi$ are eigenstates of the analog Hamiltonian, the correct $\phi_{0,i}$ is determined by minimizing the population oscillations. {\bf d,} $\{\omega_{i}\}$ are calibrated by performing adiabatic ground state preparation with an initial staggered field and a slow ($25/g_{\mathrm{m}}$) ramp, and measuring the $\langle XX\rangle$ correlations a time $t$ after the ramp. {\bf e,} Top: $\langle XX\rangle$ after time $t$ when applying no corrective phase after the analog evolution. Since the low-energy final state is known to have long-range correlations, the observed oscillations can be fit to extract the time-dependent part of the corrective phase after the analog pulse. Bottom: $\langle XX\rangle$ after time $t$ when applying the corrective phase found from fitting the oscillations. The near-constant value indicates a successful correction. {\bf f,} $\{\phi_{1,i}\}$ are calibrated by preparing an initial dimer state, performing the same circuit as in the experiment with corrective pre-analog phases $\{\phi_{0,i}\}$ and partial post-analog phases $\{\omega_i t\}$, applying a variable phase $\phi$ to one qubit in each pair, and measuring the $\langle XX\rangle$ correlations a time $t$ after the ramp. {\bf g,} Top: $\langle XX\rangle$ after time $t$. Since the state is known to be the triplet state, the correct $\phi_1$ is found from maximizing $\langle XX\rangle$ correlations. Bottom: As an complementary technique, one can prepare the singlet state instead and find the $\phi$ that minimizes variations in $\langle XX\rangle$ correlations.}  
    \label{fig:SI4}
\end{figure*}

In experiments where we prepare an entangled initial state, the frequency trajectories of the qubits lead to phase accumulation that must be characterized and corrected via phase-gates, both before and after the analog evolution (Fig.~\ref{fig:SI4}a). Specifically, in the frame that rotates at the interaction frequency, the qubits in each dimer pair precess relative to each other before they reach the interaction frequency. Hence, a phase rotation $\phi_{0,i}$ must be applied to every qubit before turning on the analog Hamiltonian to ensure that the dimer pairs have the desired phase difference when the coupling is turned on. Second, in the idle frame (in which we perform the final measurements) the qubits are precessing relative to each other while on resonance. Hence, a final phase correction $\phi_{1,i}+\omega_it$ (where $t$ is the analog evolution time) must also be applied to every qubit before measurements. Importantly, these corrections are very sensitive to timing and dispersive shifts: before the analog evolution, a timing delay in dimer generation of only 150 ps corresponds to a 0.1 rad change in $\phi_0$ for an idle frequency difference of $100$ MHz. Furthermore, during the idle evolution, a $0.1\%$ (80 kHz) change in dispersive shift leads to a 0.1 rad change in the final phase after 200 ns of analog evolution. Hence, standard calibration techniques, such as single-qubit Ramsey spectroscopy, in which the configuration is sufficiently different from that in the actual experiment, are not accurate enough. We therefore employ a set of three calibration techniques for $\phi_{0,i}$, $\phi_{1,i}$ and $\omega_i$ that are designed to represent the configuration used in the actual experiment as well as possible:

\noindent{\boldmath $\phi_{0,i}$}: To calibrate $\phi_{0,i}$, we make use of the fact that the dimer state is only an eigenstate of the coupling Hamiltonian when the phase difference of the qubits is $0$ or $\pi$. Hence, we sweep the phase difference and measure the population oscillations between the qubits with time. The correct phase compensation is the one that minimizes the amplitude of the population oscillations. We note two important points about this calibration step: first, since the measurements are in the $Z$-basis, they do not depend on the calibration of $\phi_{1,i}$ and $\omega_i$. Second, since the phase calibrated in this step is accumulated before the couplers are turned on, it is not affected by dispersive shifts. It is therefore not a problem that neighboring couplers are turned off during this particular step. 

\noindent{\boldmath $\omega_{i}$}: As mentioned previously, the calibration of $\omega_{i}$ is very sensitive to dispersive shifts and must therefore be performed in the exact same configuration as the actual experiment. We achieve this by performing the Kibble-Zurek experiment (ramp from Neel state in staggered field) with a slow ramp and leaving the analog Hamiltonian on for a variable time (Fig.~\ref{fig:SI4}d). The resultant state exhibits long-range $XX+YY$-correlations, and the effect of the phase accumulation in the idle frame is to cause oscillations in the correlator between each pair $i$ and $j$ with a frequency $\omega_{i}-\omega_{j}$ (Fig.~\ref{fig:SI4}e). Hence, by measuring the frequency of oscillations of all the correlators, the full set of $\{\omega_{i}\}$ can be determined. The key advantage of this calibration measurement is that all the couplers are turned on, so that the dispersive shifts are the same as in the actual dimer experiment. However, the initial part of the Kibble-Zurek circuit --- including the initial staggered field and the slow ramp of the couplers --- is different, so the time-independent part of the phase correction, $\phi_{1,i}$, must be calibrated separately.

\noindent{\boldmath $\phi_{1,i}$}: Finally, to determine $\phi_{1,i}$, we take advantage of energy conservation. Specifically, we perform the dimer experiment with single dimers while sweeping their final phase difference (Fig.~\ref{fig:SI4}f). Only the correct phase compensation leads to $\langle X_1X_2\rangle=1$ and conserved energy, as can be see in Fig.~\ref{fig:SI4}g. While the dispersive shifts from neighboring couplers impact the time-dependent part of the final phase $\omega_it$ and thus had to be included in the previous step, they do not have this effect on $\phi_{1,i}$ and can therefore be excluded here.

Finally, we note that for experiments not involving entangled initial states (Figs. 3 and 4 in the main text), only the step for calibration of $\{\omega_i\}$ outlined above is required.

\subsection{Readout correction and postselection schemes}
\subsubsection{Bell measurements}

When measuring $\langle XX+YY\rangle$ correlators using standard single-qubit measurements, we cannot simultaneously get information about the number of photons measured on the pair of qubits, preventing us from postselecting our data on photon conservation. To get around this for nearest neighbor pairs, we change our measurement basis by applying an entangling gate given by the unitary, 
\begin{equation*}
    \begin{bmatrix}
        1 & 0 & 0 & 0 \\
        0 & 1/\sqrt{2} & -1/\sqrt{2} & 0 \\
        0 & 1/\sqrt{2} & 1/\sqrt{2} & 0 \\
        0 & 0 & 0 & 1
    \end{bmatrix},
\end{equation*}
to each pair to get the conversion shown in Table \ref{tab:bell}. From these measurements, we can deduce both the nearest neighbor correlators and the number of photons present. We use this technique to process the data labeled `Bell' in Figure~3b of the main text. We find good alignment between direct measurements of the correlators and the inferred correlators from the Bell measurements.

\subsubsection{Bell measurements with readout corrections}
Typically, we correct readout errors by inverting the error channel. In the case where readout errors are uncorrelated, we can simply characterize the matrix $\beta$ for each qubit
\begin{equation*}
    \beta = \begin{bmatrix} p_{(0|0)} & p_{(0|1)} \\ p_{(1|0)} & p_{(1|1)} \end{bmatrix}
\end{equation*}
where $p_{(i|j)}$ is the probability of measuring a state $\ket{i}$ given that $\ket{j}$ was prepared \cite{ro_corr}. In the case where readout errors are correlated for pairs, we can similarly characterize a matrix $\gamma$ for each pair
\begin{equation*}
    \gamma = \begin{bmatrix} p_{(00|00)} & p_{(00|01)} & p_{(00|10)} & p_{(00|11)} \\ 
    p_{(01|00)} & p_{(01|01)} & p_{(01|10)} & p_{(01|11)} \\
    p_{(10|00)} & p_{(10|01)} & p_{(10|10)} & p_{(10|11)} \\
    p_{(11|00)} & p_{(11|01)} & p_{(11|01)} & p_{(11|11)} \\
    \end{bmatrix}
\end{equation*}
where $p_{(ij|ab)}$ is the probability of measuring a state $\ket{ij}$ given that $\ket{ab}$ was prepared. Inverting these matrices and applying them to their respective state vectors allow us to reconstruct the state vector unaffected by their readout errors. 
\begin{table}[]
    \centering
    \begin{tabular}{c|c|c|c|c}
        Prepared & Post gate & $XX$ & $YY$ & $N_{\text{photons}}$ \\ \hline
        $\ket{00}$ & $\ket{00}$ & 0 & 0 & 0 \\
        $\ket{01} - \ket{10}$ & $\ket{01}$ & -1 & -1 & 1 \\
        $\ket{01} + \ket{10}$ & $\ket{10}$ & 1 & 1 & 1 \\
        $\ket{11}$ & $\ket{11}$ & 0 & 0 &2 \\
    \end{tabular}
    \caption{Summary of the following quantities for each prepared state: the states after converting to the Bell basis, $XX$ and $YY$ correlators, and the number of photons in the pair.}
    \label{tab:bell}
\end{table}

In a case where we want to both correct for readout errors and postselect our data, we cannot apply the readout correction on the state vector reconstructed from the postselected bitstrings since this would overcorrect for $p_{(0|1)}$ type errors. We also cannot invert the matrices and apply them to the state vectors before the postselection process since we need access to the individual bitstrings to postselect on photon number conservation. Instead, we use a Markov-like process in which we consider each individual bitstring, and flip spins according to the probabilities inferred from our $\beta$ or $\gamma$ matrices. We then postselect the individual bitstrings on the criteria of photon conservation and, finally, compute the quantity of interest. 

\begin{figure}[!htbp]
    \centering
    \includegraphics[width=\columnwidth]{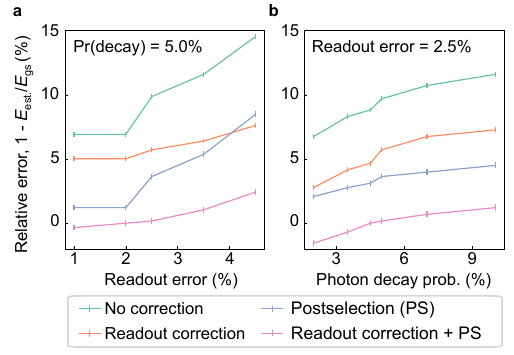}
    \caption{{\bf Correction for readout error and photon decay.}  Performance of various readout and photon decay correction techniques as a function of {\bf a,} readout error {\bf b,} and photon decay probability. The performance is measured as the relative error between the estimated energy ($E_{\mathrm{est.}}$) and the actual ground state energy ($E_{\mathrm{gs}}$). We find that the combined technique (pink) achieves the lowest relative error.}
    \label{fig:SI_RO_corr}
\end{figure}

To confirm the validity of this method, we classically simulate the ground state of the XY-model with 20 qubits, introduce noise to the system, and use the above protocol to correct for the $T_1$ and readout errors. We compute the energies of the system after various correction schemes and compare to the noiseless value. The results from these simulations are found in Figure~\ref{fig:SI_RO_corr}a,b, where we evaluate the performance as a function of the readout error and probability of photon decay, respectively. We find that postselection performs somewhat better than readout correction for sufficiently low readout errors; however, most importantly, the combined technique described above achieves the lowest error. 
We note that the relative error of the method has a non-trivial dependence on the $T_1$ errors, but that the combined protocol outperforms the other error mitigation techniques and bring us closest to the energy of the noiseless state across the full parameter space we explore. In the experiment, we have readout errors in the range $1$-$4\%$ and a probability of photon decay of $3$-$6\%$ for ramp times of $200$-$500$ ns.

\subsection{\texorpdfstring{Comparison of $\langle XX\rangle$ and $\langle YY\rangle$}{Comparison of <XX> and <YY>}}
\begin{figure}[!htbp]
    \centering
    \includegraphics[width=1\columnwidth]{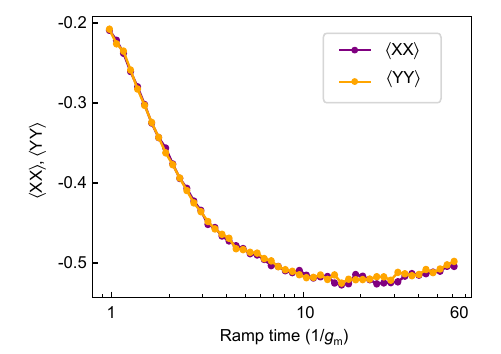}
    \caption{{\bf Comparison of $XX$- and $YY$-correlations.} Ramp time dependence of $\langle XX\rangle$ and $\langle YY\rangle$ averaged over all nearest-neighbor pairs. The two are found to be equally strong, consistent with $U(1)$-symmetry.}   
    \label{fig:SI_xxvsyy}
\end{figure}

The final states produced after the ramp procedures in Figs. 3 and 4 in the main text are expected to be $U(1)$-symmetric, and thus have equally strong $XX$- and $YY$-correlations. We here check this by comparing $\langle XX\rangle$ and $\langle YY\rangle$ averaged over all nearest-neighbor qubit pairs across a range of ramp times (Fig.~\ref{fig:SI_xxvsyy}), and indeed find that the two are equal.

\subsection{Measurements of energy density fluctuations}
\begin{figure}[!htbp]
    \centering
    \includegraphics[width=1\columnwidth]{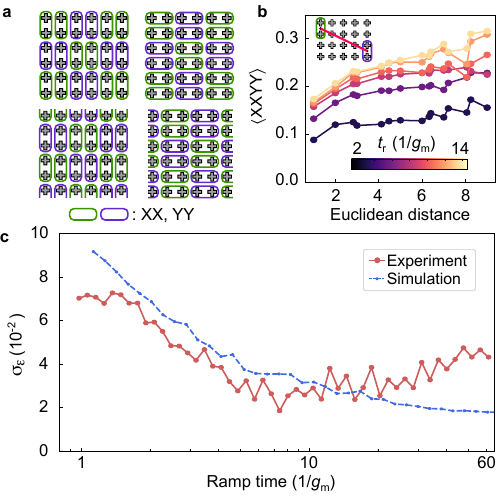}
    \caption{{\bf Energy density fluctuations. a,} In addition to $\{X_i\}$, $\{Y_i\}$ and $\{Z_i\}$, we measure 8 periodic patterns of $XX$ and $YY$ to find $\sigma_{\varepsilon}$. {\bf b,} $\langle XXYY \rangle$ has a relatively simple dependence on Euclidean distance (data from measurements shown in {\bf a}), which can be interpolated to find the remaining terms. {\bf c,} Energy density fluctuations, $\sigma_{\epsilon}$, displaying good agreement between experiment (red) and simulation (blue); however, at long ramp times, decoherence causes higher fluctuations in the experimental case.  
    }   
    \label{fig:SI_dE}
\end{figure}

In the main text, we use measurements of 2- and 4-qubit correlators to reconstruct the energy density fluctuations, $\sigma_{\varepsilon}=(n_{\mathrm{B}}g_{\mathrm{m}})^{-1}\sqrt{\langle H_{XY}^2\rangle-\langle H_{XY}\rangle^2}$, with:
\begin{align}
    \frac{H_{XY}^2}{g^2_{\mathrm{m}}} = \left(\sum_{\langle i,j\rangle} (X_iX_j+Y_iY_j)/2\right)^2= \sum_{\langle i,j\rangle} (1-Z_iZ_j)/2\\
     \nonumber +\sum_{\langle i,j\rangle}\sum_{\langle m,n\rangle}(X_iX_jX_mX_n+Y_iY_jY_mY_n+X_iX_jY_mY_n+ \\ + \nonumber Y_iY_jX_mX_n)/4
    +\sum_{\langle i,j\rangle,\langle j,k\rangle } (X_iX_k+Y_iY_k)/2,
\end{align}
where $\langle i,j\rangle$, $\langle j,k\rangle$ and $\langle m,n\rangle$ are nearest neighbor pairs and $i,j,k,m,n$ are distinct (note that $j$ is included in the last sum to count the number of length-2 paths from $i$ to $k$). Importantly, almost all of these terms can be reconstructed from just three different sets of measurements, namely $\{X_i\}$, $\{Y_i\}$ and $\{Z_i\}$, except the 4-qubit correlators involving both $X$ and $Y$. In order to determine these, we measure 8 periodic patterns of $X$ and $Y$ shown in Fig.~\ref{fig:SI_dE}a, and take advantage of the isotropicity of our system. As shown in Fig.~\ref{fig:SI_dE}b, the 4-qubit correlators that involve both $X$ and $Y$ show a clear trend with the Euclidean distance between the centers of mass of the two involved nearest-neighbor pairs ($i,j$) and ($m,n$), and we therefore interpolate the data obtained from these 8 sets of measurements to find the remaining terms. 
Determining $\sigma_{\varepsilon}$ with good relative accuracy is challenging, due to the very small relative difference between $\langle H_{XY}\rangle ^2$ and $\langle H_{XY}^2\rangle$. Nevertheless, we find that our technique works well, and obtain relatively good agreement with matrix product state (MPS) simulations (Fig.~\ref{fig:SI_dE}c).

\subsection{Numerical finite-size scaling analysis}
\begin{figure*}[t!]
   \centering
   \includegraphics[width=1.0\textwidth]{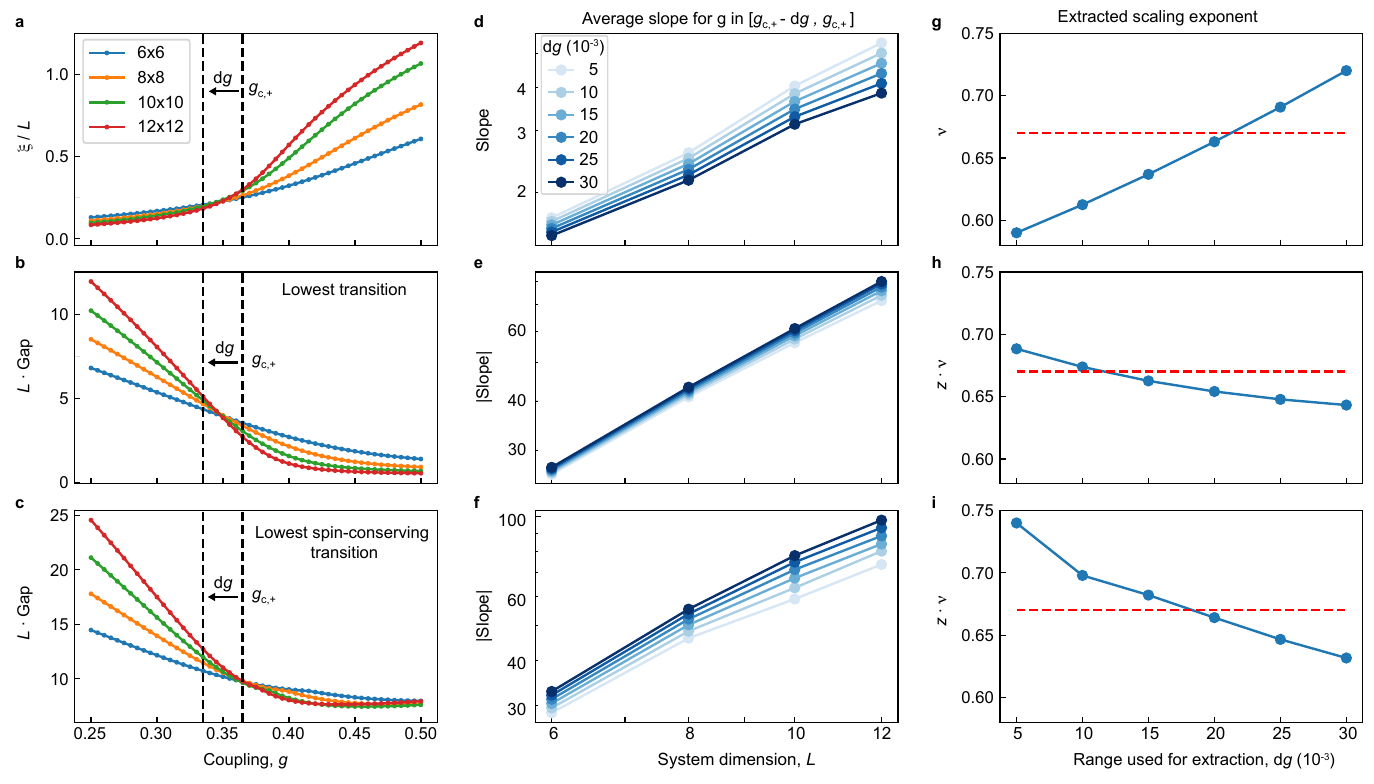}
   \caption{{\bf Numerical extraction of scaling exponents through finite-size scaling analysis. a,} Correlation length of ground state normalized by system dimension $L$, determined from MPS simulations and plotted versus coupling, $g$ (see text for further details). Finite-size crossing is observed near $g_c=0.35$. Black dashed vertical lines indicate extrema of ranges used for slope extraction. {\bf b,c} Same as {\bf a}, but for gap multiplied by system dimension. We consider both the overall lowest transition ({\bf b}) and the smallest spin-conserving one ({\bf c}). {\bf d-f,} Extracted slopes from {\bf a-c}, respectively, for ranges of various widths, and plotted as a function of system dimension, $L$. The slopes grow near-algebraically with $L$, as expected. {\bf g-i,} Extracted scaling exponents from {\bf d-f}, respectively. In all cases, the extracted exponents are consistent with the predicted $z=1$, $\nu=0.67$ (red dashed line).}
   \label{fig:SI_scaling}
\end{figure*}
We here perform finite-size scaling analysis to evaluate the critical scaling exponents $\nu$ and $z$ near the critical point in the XY-model studied in our work. We perform DMRG simulations on square-shaped clusters of $L\times L$ sites to determine the correlation length of the ground state (Fig.~\ref{fig:SI_scaling}a), as well as the gap size, $\Delta$, of both the lowest transition ($S_z=0\rightarrow S_z=1$; Fig.~\ref{fig:SI_scaling}b) and the lowest spin-conserving transition ($S_z=0\rightarrow S_z=0$; Fig.~\ref{fig:SI_scaling}c), while sweeping the staggered field ($h$) from 1 to 0 and the coupling ($g$) from 0 to 1. We use a bond dimension $\chi=2048$ and system sizes $L\times L$ with $L\in \{6,8,10,12\}$. 
Near the critical point, it is expected that $\xi/L=F_{\chi}((g-g_c)L^{1/\nu})$ and $L\Delta=F_{\Delta}((g-g_c)L^{1/(z\nu)})$. Motivated by this, we plot $\xi/L$ and $L\Delta$, and observe finite-size crossing near $g_c=0.35$. As can be seen from the expressions above, the slope near the critical point is expected to scale as $L^{1/\nu}$ and $L^{1/(z\nu)}$ for the correlation length and gaps, respectively. Hence, to evaluate the scaling exponents while also accounting for variations in slope near the critical point, we extract the slope in ranges of varying width $dg$ from $g_{c,+}-dg$ to $g_{c,+}$ with $g_{c,+}=0.365$ (Figs.~\ref{fig:SI_scaling}d-f). As theoretically expected, we find that the slopes increase near-algebraically with $L$, from which we extract scaling exponents shown in Figs.~\ref{fig:SI_scaling}g-i. For all three cases, we find that the extracted exponents are consistent with the expected $\nu=0.67$, $z=1$.

\subsection{Empirical estimation of self-XEB}
\label{appendix:selfXEB}
\subsubsection{Ideal case}
\begin{figure}[h!]
    \centering
    \includegraphics[width=1.0\columnwidth]{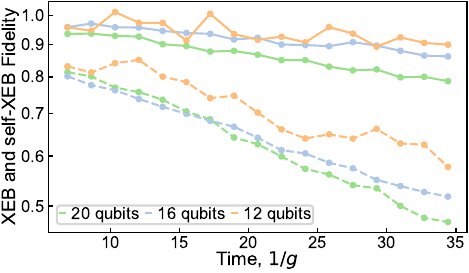}
    \caption{{\bf Incoherence fidelity from self-XEB.} Solid lines show the estimate of the fidelity $\Phi$ describing the global depolarizing channel Eq.\,\eqref{eq:gpc} using the expression Eq.\,\eqref{eq:F_incoherent}. The dashed lines show the XEB with the experimental data used in the main text.}
    \label{fig:SI_self_XEB_Fidelity}
\end{figure}
In the main text, we measure self-XEB by constructing an empirical bitstring distribution $p_{\text{exp.}}(s_i) = \frac{M_i}{M},$ where $M_i$ is the total number of times a bitstring $s_i$ was sampled from a quantum device (after postselecting bitstrings with the correct number of excitations), and $M$ is the total number of postselected bitstrings. An empirical estimate for the self-XEB reads
\begin{gather}
    \text{self-XEB}_{\text{est.}} = D \sum\limits_{s_i} \frac{M_i^2}{M^2} - 1,
\end{gather}
and $M_i$ follows the binomial distribution $P(M_i = x) = {M \choose x} p_i^x (1 - p_i)^{M - x}$, where $p_i$ is the quantum probability of sampling a bitstring $s_i$. The second moment of the binomial distribution reads $\displaystyle \mathop{\mathbb{E}} M_i^2 = M^2 p_i^2 + M p_i (1 - p_i)$, therefore
\begin{gather}
    \mathop{\mathbb{E}} \text{self-XEB}_{\text{est.}} = \text{self-XEB}_{\text{true}} + \frac{D}{M} \sum\limits_{s_i} p_i (1 - p_i) = \\ \nonumber = \text{self-XEB}_{\text{true}} + \frac{D}{M} - \frac{1}{M} \left(\text{self-XEB}_{\text{true}} + 1 \right) = \\ \nonumber = \left(1 - \frac{1}{M}\right) \times \text{self-XEB}_{\text{true}} + \frac{D - 1}{M},
\end{gather}
which means that $\text{self-XEB}_{\text{est.}}$ is a biased estimator with the relative bias of $1/M$ and the absolute bias of $(D - 1) / M$. To account for this, in the main text, we estimate self-XEB using an unbiased estimator 
\begin{gather}
\label{eq:XEB_unbiased}
    \text{self-XEB}_{\text{est., unbiased}} = \frac{\text{self-XEB}_{\text{est.}} }{1 - 1 / M} - \frac{D - 1}{M - 1}. 
\end{gather}

\subsubsection{Depolarizing channel}
Study of the bitstring distribution provides information about incoherent errors under some assumptions on the noise type~\cite{boixo_2018}. For simplicity, we consider a global depolarizing channel, which modifies bitstring probabilities as 
\begin{gather}
\label{eq:gpc}
    \tilde p_i = \Phi p_i + (1 - \Phi)/D,    
\end{gather}
where $p_i$ is a bitstring probability in the ideal case, and $\Phi$ is the fidelity of the system with respect to ideal, closed-system evolution. This approximation yields
\begin{gather}
    \text{self-XEB}_{\text{noisy}} = D \sum\limits_i \tilde p_i^2 - 1 = \Phi^2 \times \text{self-XEB}_{\text{ideal}}.
\end{gather}
Therefore, incoherent fidelity $\Phi$ may be estimated as
\begin{gather}
\label{eq:F_incoherent}
    \Phi = \sqrt{\text{self-XEB}_{\text{noisy}} / \text{self-XEB}_{\text{ideal}}},
\end{gather}
where $\text{self-XEB}_{\text{ideal}} = D \sum_i p_i^2 - 1$ is estimated using an ideal classical simulation, and $\text{self-XEB}_{\text{noisy}}$ is estimated empirically from experimental data using the unbiased estimator Eq.\,\eqref{eq:XEB_unbiased}. With this benchmark, we get an independent estimate of the system's incoherent errors. The result is shown in Fig.\,\ref{fig:SI_self_XEB_Fidelity}. We conclude that the error rates quoted in the main text are mainly dominated by coherent sources. We note  that although the decoherence processes in this experiment  are mainly due to dephasing (since postselection eliminates decay processes), we expect the simplified depolarizing-channel model to correctly capture the magnitude of incoherent processes.

\subsection{Exact state-vector simulation}
\label{appendix:exactsim}
The time evolution performed in the experiment consists of a short time-dependent on-ramp, a long  time-independent ``plateau'', and a short time-dependent off-ramp. To simulate the time-dependent ramps, we solve the time-dependent Schrödinger equation using the \texttt{Runge-Kutta-45} algorithm.

For stationary evolution, we employ the Chebyshev polynomials approach~\cite{HERNANDEZ2001433,Yuan_2010}. For $x \in [-1, 1]$ and any real $\tau$, an exponential can be decomposed as 
\begin{gather}
    \label{eq:cheb_exponent}
    e^{-i x \tau} = \sum\limits_{m=0}^{+\infty} \alpha_m (-i)^m J_m(\tau) T_m(x),
\end{gather}
where $\alpha_0 = 1$, $\alpha_{m} = 2$ for $m > 0$, $J_m(\tau)$ is a Bessel function of the first kind of the $m$-th order, and $T_m(x) = \cos[m \arccos(x)]$ is the $m$-th Chebyshev polynomial. These polynomials obey the following recurrence relation:

\begin{gather}
    T_{m + 1}(x) + T_{m - 1}(x) = 2\,x\,T_m(x), \\ \nonumber \; T_0(x) = 1, \; T_1(x) = x.
\end{gather}

To time-evolve a wave function $|\psi(0)\rangle$ for a time $t$ with a Hamiltonian $\hat{H}$, we need to apply the Chebyshev decomposition Eq.\,\eqref{eq:cheb_exponent} to the matrix exponential $\exp\left(-i t \hat{H}\right)$. To this end, we introduce a rescaled Hamiltonian

\begin{gather}
    \hat{h} = \frac{1}{E_{\text{max}} - E_{\text{min}}} \hat{H} - \frac{E_{\text{max}} + E_{\text{min}}}{2 (E_{\text{max}} - E_{\text{min}})} \hat{I},
\end{gather}
where $\hat{I}$ is the identity operator, and $E_{\text{min}},\,E_{\text{max}}$ are the minimum and maximum eigenvalues. The resulting rescaled Hamiltonian $\hat{h}$ has its spectrum within $[-1, +1]$, as the decomposition Eq.\,\ref{eq:cheb_exponent} requires. In practice, it is sufficient to only set an upper bound on the bandwidth $W = (E_{\text{max}} - E_{\text{min}})$ to ensure and all eigenspectrum lies within $[-1, +1]$. Applying a corresponding rescaling of the evolution time $\tau = t \times (E_{\text{max}} - E_{\text{min}})$, we obtain

\begin{gather}
    |\psi(t)\rangle = J_0(\tau) |\psi^{\hat{h}}_0\rangle + 2 \sum\limits_{m = 1}^{+\infty} J_m(\tau) |\psi^{\hat{h}}_m\rangle,
\end{gather}
where we defined the Chebyshev partons:
\begin{gather}
\label{eq:cheb_partons}
    |\psi^{\hat{h}}_{m + 1}\rangle = (-i)^{m + 1}T_{m + 1}(\hat{h}) |\psi(t = 0)\rangle = \\
    \nonumber -2 i \hat{h} |\psi^{\hat{h}}_{m}\rangle + |\psi^{\hat{h}}_{m - 1}\rangle,\\
    \nonumber |\psi^{\hat{h}}_{0}\rangle = T_0(\hat{h}) |\psi(t = 0)\rangle = |\psi(t = 0)\rangle,\\
    \nonumber |\psi^{\hat{h}}_{1}\rangle = (-i) T_1(\hat{h}) |\psi(t = 0)\rangle = -i \hat{h} |\psi(t = 0)\rangle.
\end{gather}
For $\|\hat{h}\| < 1,$ all Chebyshev polynomials are bounded, $\|T_m(\hat{h})\| < 1$, which guarantees stability and convergence of the algorithm. To obtain the required number of matrix-vector operations, we consider large-$m$ asymptotics of the Bessel functions for $\tau \ll \sqrt{m + 1}$:

\begin{gather}
    J_{m}(\tau) \sim \frac{1}{\Gamma(m + 1)} \left(\frac{\tau}{2}\right)^{m} \sim \frac{1}{\sqrt{2 \pi m}} \left( \frac{e\tau}{2 m}\right)^m,
\end{gather}
which makes this Bessel function reach a maximum at $m = m^* / e = \tau / 2$ and then decay super-exponentially with $m$ for $m > m^* = e \tau / 2 = (e / 2) \times (E_{\text{max}} - E_{\text{min}}) t.$ Thus, the threshold $m^*$ defines the typical number of matrix-vector actions that are necessary to time-evolve a wave function for time $t$. Therefore, we see that the Chebyshev time evolution algorithm complexity is linear in the evolution time, the Hamiltonian bandwidth, and Hilbert space dimension, $\mathcal{O}(D t W)$. 

The recurrence relation Eq.\,\ref{eq:cheb_partons} requires storing 4 vectors. We stop iterating when the norm of the time-evolved state satisfies $|\||\psi(t)\rangle\| - 1| < 10^{-10}.$ Finally, given the time stamps ${t_0,\,t_1,\,\ldots}$ used to produce experimental bitstring samples, we time-evolve our wave function consecutively between these time stamps, setting $t = t_{i + 1} - t_{i}$ and starting from a previously obtained wave function. 

To perform these simulations, we use the \texttt{lattice-symmetries} package~\cite{Westerhout_2021}, which utilizes excitation number conservation and matrix-less matrix-vector operations. A single \texttt{c3d-standard-360} node with 1.4\,Tb RAM on the Google Cloud Platform allows to exactly (up to the $10^{-10}$ stopping criterion error) time-evolve a wave function with the Hilbert space dimension of $\sim 9 \times 10^{9}$ for $t \times (E_{\text{max}} - E_{\text{min}}) = 2 \times 10^3$ in only 5 days. This Hilbert space size corresponds to 18 excitations in a 36-qubit system, or to 8 excitations in a 64-qubit system.
\begin{figure}[t!]
    \centering
    \includegraphics[width=1.0\columnwidth]{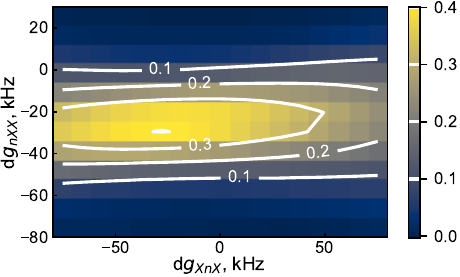}
    \caption{{\bf Optimizing the XEB fidelity.} Here, we show XEB fidelity for different values of the uniform global shifts in $g^{nXX}$ and $g^{XnX}.$ We consider the data from the main text with the system size of 25 qubits. The scan suggests an optimum shift at $\mathrm{d}g^{nXX} \approx \mathrm{d}g^{XnX} \approx -20\,\text{kHz}$.}
    \label{fig:SI_XEB_fitting}
\end{figure}

\subsection{Few-parameter XEB optimization}
In the calibration protocol used in this work, we produce terms in the effective spin-Hamiltonian $\hat{H}_{\text{s}}$ by considering sub-systems of the chip. Therefore, a small possible bias in this procedure (due to the patch Hilbert space truncation) would be replicated in the whole system and cause coherent errors~\cite{cai2023stochastic}. We ameliorate this effect by performing {\it a posteriori} XEB optimization over only two global parameters, namely uniform shifts in the couplings $g^{XnX}$ and $g^{XXn}$ defined in Eq.\,\eqref{eq:higher_order} (Fig.\,\ref{fig:SI_XEB_fitting}). All other global shifts do not improve the XEB. This two-parameter fit finds very small optimal shifts of $\mathrm{d}g^{XnX} \approx \mathrm{d}g^{nXX} \approx -20\,\text{kHz}$, which are optimal for all system sizes and occupations, thus confirming our assumption about the systematic nature of these coherent errors.

\subsection{Fidelity prediction at a given system size and time}
\label{appendix:fidelity_fitting}
\begin{figure}[h!]
    \centering
    \includegraphics[width=1.0\columnwidth]{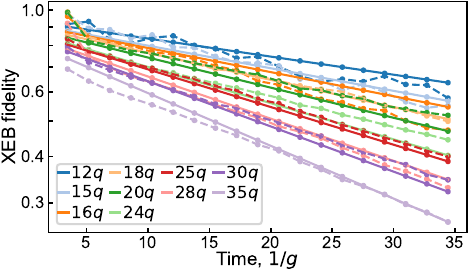}
    \caption{{\bf Fitting the XEB fidelity} Solid lines are the experimental results also shown in the main text, and the dashed lines show a global fit using Eq.\,\eqref{eq:fidelity_ansatz}.}
    \label{fig:SI_XEB_fits}
\end{figure}

In this work, time evolution consists of a fast on-ramp, time evolution with a Hamiltonian $H_s$ over a much longer time, and a fast off-ramp. We write an ansatz for XEB fidelity expected after a fixed  evolution time $t$ with $H_s$ on a system with size $N_{\text{q}}$:
\begin{gather}
    \label{eq:fidelity_ansatz}
    F(t, N_{\text{q}}) = F_0^{N_{\text{q}}} e^{-\epsilon \times N_{\text{q}} \times (t / T)},
\end{gather}
where $\epsilon$ is the per-qubit-per-cycle error, $T$ is the cycle time, and $F_0$ accounts for fidelity loss during a 6-ns on-ramp, 6-ns off-ramp, and readout. We fit the XEB fidelities obtained in the dataset from the main text with system sizes from 12 to 35 qubits, and obtain $F_0 = 0.9946,\; \epsilon = 9.4 \times 10^{-4}.$ The resulting fits are shown in Fig.\,\ref{fig:SI_XEB_fits}, and the fit yields root-mean-square-error of $2.4 \times 10^{-3}.$

\subsection{Entanglement of time-evolved states}
\label{subsec:entanglement}

\begin{figure*}[t!]
    \centering
    \includegraphics[width=\textwidth]{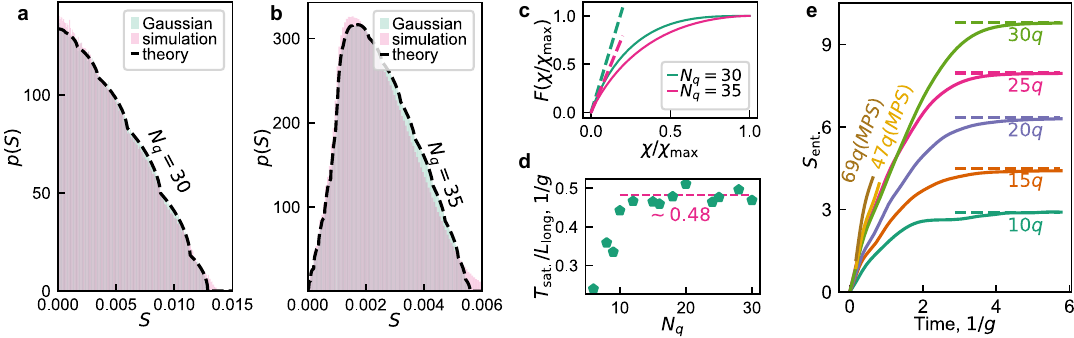}
    \caption{{\bf Entanglement properties of the simulated wave functions.} {\bf a,b,} Distribution of Schmidt values in a 30- and 35-qubit systems, where we compare the quarter-circle theory, random Gaussian ensemble, and wave functions obtained in the simulation of the effective spin Hamiltonian. {\bf c,} State fidelity as a function of the bond dimension cutoff. The dashed lines show linear slope $F \leqslant \zeta(N_{\text{q}}) (\chi / \chi_{\text{max}}) \approx 4 \sqrt{2} (\chi / \chi_{\text{max}})$. {\bf d,} The time it takes for the system to build 90\,\% of the maximum entanglement at given system size, divided by the longest dimension of the rectangle $L_{\text{long}}.$ The linear fit at $\sim 0.48 \times (1/g)$ indicates convergence $t_{\text{sat.}} \sim (1 / 2 g) \times L_{\text{long}}.$ {\bf e,} Entanglement entropy as a function of time in classical ideal simulations for various system sizes using state-vector and MPS approaches. The MPS simulations are stopped upon saturation of the bond dimension. The dashed lines show entanglement entropy of typical half-filled states estimated using Eq.\,\eqref{eq:Sent_U1}.}
    \label{fig:SI_entanglement}
\end{figure*}

In this section, we study the growth of entanglement in our system, and the evolution of its Schmidt spectrum. Most of the computations are performed using exact state-vectors, and some simulations employ matrix-product states. Most importantly, we analyze the effects of particle-number conservation on the entanglement properties.

\subsubsection{Schmidt values distribution}

For all the (rectangular) geometries, we compute entanglement entropy between two parts of the $N$-qubit system of sizes $L = \lfloor N / 2 \rfloor$ and $R = N - L$. The left part of the system includes all sites with $x + y \times L_x < L$, where $0 \leqslant x < L_x$, $0 \leqslant y < L_y$, and $L_x \leqslant L_y$, i.\,e., the system is cut into two halves along the shortest direction.

Consider a system with $M$ particles and $N = L + R$ sites, with a wave function $|\psi\rangle$ written in the full $2^{N}$-element basis. A matrix element of the reduced density matrix (RDM) of the left-hand side reads
\begin{gather}
    \left(\hat \rho_L\right)_{ij} = \left(\Tr_R\, \left[|\psi \rangle \langle \psi | \right]\right)_{ij} = \sum\limits_{r=0}^{2^{R} - 1} \psi^*_{i, r} \psi_{j, r},
\end{gather}
where $i (j)$ and $r$ enumerate basis states in the $L$ or $R$ subsystem, respectively. Since the total number of excitations $M$ is fixed, the matrix elements are non-zero only for $(i, j)$ satisfying 
\begin{gather}
    \text{Ham}[i] + \text{Ham}[r] = \text{Ham}[j] + \text{Ham}[r] = M, 
\end{gather}
where $\text{Ham}[k]$ counts the number of excitations in a bitstring $k$ (Hamming weight). Therefore, $\text{Ham}[i] = \text{Ham}[j] = M - \text{Ham}[r]$, and the reduced total density matrix has a block-diagonal form
\begin{gather}
    \label{eq:factorization}
    \hat{\rho}_{L} = \bigotimes\limits_{M_L = 0}^{\text{min}(L, M)} \hat{\rho}_{L, M_L},
\end{gather}
where $0 \leqslant M_L \leqslant \text{min}(L, M)$ is the number of excitations in the left-hand side of the system. At half-filling, the total number of Schmidt values is $2^L$. To obtain the Schmidt spectrum, we perform eigendecomposition of all partial density matrices $\hat{\rho}_{L, M_L}$.

We now consider typical distributions of the Schmidt values in the $U(1)$--conserving case at infinite temperature. In a generic case (without particle number conservation), the RDM is irreducible. If the system is divided into the left and right parts with $L$ and $R$ sites and the effective Hilbert space sizes $D_L, D_R$ ($D_L D_R = 2^N)$, the Schmidt values follow the generalized quarter-circle distribution~\cite{marcenko}
\begin{gather}
    p(S) = \frac{D}{\pi}\frac{\sqrt{(\lambda_+^2 / D_L - S^2) (S^2 - \lambda_-^2 / D_L)}}{\lambda S},
\end{gather}
where $\lambda = D_L / D_R$, $\lambda_{\pm} = 1 \pm \sqrt{\lambda}$, and $\lambda_- / \sqrt{D_L} \leqslant S \leqslant \lambda_+ / \sqrt{D_L}$. 

In contrast, in the $U(1)$--symmetric case, the Schmidt values follow the quarter-circle distribution independently in each $M_L$--block of the RDM~\cite{cheng2023typical}. The $U(1)$--constrained quarter-circle distribution of the Schmidt values $S_i$ reads

\begin{gather}
\displaystyle
    p(S) =  \sum\limits_{M_L=0}^{L} \begin{cases}
        0,\; S \notin [\lambda_-^{M_L}, \lambda_+^{M_L}], \\
        p_{M_L}(S), \; S \in [\lambda_-^{M_L}, \lambda_+^{M_L}],\label{eq:Schmidt_formula}
    \end{cases}, \text{with} \\
    \nonumber p_{M_L(S)} = \frac{\displaystyle D}{\displaystyle S \pi } \times \sqrt{\left[ \left(\lambda_+^{M_L}\right)^2 - S^2\right] \left[S^2 - \left(\lambda_-^{M_L}\right)^2\right]},
\end{gather}
where $D = {N \choose M}$ is the full Hilbert space dimension, $D_L^{M_L} = {L \choose M_L}$ is the number of ways to put $M_L$ particles in the left part, $D_R^{M_L} = {R \choose M - M_L}$ is the number of ways to put $(M - M_L)$ particles in the right part, and $\lambda_{\pm}^{M_L} = (1 \pm \sqrt{\lambda^{M_L}}) \sqrt{D_R^{M_L} / D}$ with $\lambda^{M_L} = D_L^{M_L} / D_R^{M_L}$.

In Fig.\,\ref{fig:SI_entanglement}a,b, we show the Schmidt values distribution of the 30 and 35-qubit systems with 15 and 18 photons, respectively. The distributions are obtained (i) using the exact simulation of the system Hamiltonian, (ii) from Schmidt-decomposing a wave function with random complex Gaussian entries $\psi_i \sim \mathcal{N}(0, 1) + i \mathcal{N}(0, 1)$, and (iii) applying the equation Eq.\,\eqref{eq:Schmidt_formula}, which accurately describes the Schmidt value distributions. We conclude that the time evolution with the Hamiltonian $H_{\text{s}}$ generates typical states, in the sense of $U(1)$--constrained quarter-circle Schmidt value distribution.

The complexity of a MPS encoding of time-evolved states is determined by the fraction $\chi / \chi_{\text{max}}$, where $\chi_{\text{max}}$ is the maximum possible number of Schmidt values given Hilbert space sizes, required to represent a wave function with a given fidelity $\mathcal{F}$. In Fig.\,\ref{fig:SI_entanglement}c, integrating the distribution Eq.\,\eqref{eq:Schmidt_formula}, we plot $\mathcal{F}(\chi / \chi_{\text{max}})$. In the half-filled case, the derivative at zero is given by
\begin{gather}
    \label{eq:F_chi_bound}
    \mathcal{F} \leqslant 4 \frac{\chi}{\chi_{\text{max}}} \frac{2^L \sqrt{{L \choose L / 2} {R \choose R / 2}}}{{N \choose N / 2}} = 4 \frac{\chi}{\chi_{\text{max}}} \left(\sqrt{2} + \mathcal{O}(1/N)\right),
\end{gather}
which differs by a factor $\sqrt{2}$ from the generic case with no particle conservation~\cite{morvan2023phase}. This formula will be used below to derive bounds on complexity of the experimentally realized states. 

\subsubsection{Entanglement entropy}
In Fig.\,\ref{fig:SI_entanglement}d,e we consider the entanglement entropy growth with time at various system sizes. The dashed lines indicate the maximum entanglement entropy at a given system size at half-filling with $U(1)$ conservation. Using Eq.\,\eqref{eq:Schmidt_formula}, these bounds can be computed as~\cite{lin2022exact}:
\begin{gather}
    S_{\text{ent.}} = \sum\limits_{M_L=0}^L \frac{D_R^{M_L} D_L^{M_L}}{D} \left[\log \left(\frac{D}{D_R^{M_L}}\right) - \frac{\lambda^{M_L}}{2} \right].
\end{gather}
Assuming $L = R$ and the half-filled case with even $M = (L + R )/ 2$, it can be further simplified (replacing summation over $M_L$ with a Gaussian integral):
\begin{gather}
    \label{eq:Sent_U1}
    S_{\text{ent.}}^{U(1)} = \frac{N}{2} \log 2 - \frac{1}{2} \log 2 - \frac{1}{4} + \mathcal{O}(1/N).
\end{gather}
This expression is different from $S_{\text{ent.}}^{\text{generic}} = \log \sqrt{D} - 1/2$ $= (N/2)\log 2 - 1/2 = S_{\text{ent.}}^{U(1)} + 0.097$, obtained for an irreducible RDM. Fig.\,\ref{fig:SI_entanglement}e shows that the entanglement entropy of the simulated wave functions approaches $S_{\text{ent.}}^{U(1)}$ at long times.

Fig.\,\ref{fig:SI_entanglement}d shows the ratio of the time it takes the system to build 90\,\% of the maximum entanglement entropy, $t_{\text{sat.}}$, to the longest direction of the rectangle $L_{\text{long}}.$ Since the maximum entropy is proportional to the total system volume, $\text{max}\,S_{\text{ent.}} \propto L_{\text{long}} L_{\text{short}}$, and the rate of entanglement generation at early times is proportional to the length of the shortest cut, $dS_{\text{ent.}} / dt \propto L_{\text{short}}$, we expect $t_{\text{sat.}} = \alpha L_{\text{long}}$. Indeed, Fig.\,\ref{fig:SI_entanglement}d shows that the ratio $t_{\text{sat.}} / L_{\text{long}}$ saturates at around $\alpha \approx (1 / 2 g)$ in the simulation with $g/(2 \pi) \approx 10\,\text{MHz}$.

\subsubsection{Log-negativity}
For noisy dynamics on a quantum device, it is customary to compute the log-negativity 
\begin{gather}
    \mathcal{E}_{\text{N}}(\psi) = \log_2 \| |\psi\rangle\langle\psi|^{T_A}\|_1,
\end{gather}
where $|\psi\rangle\langle\psi|^{T_A}$ is a partially-transposed (in a subsystem $A$) density matrix of a pure system with the wave function $|\psi\rangle.$ Here, we focus on a bipartition of a system into equal-sized regions $L$ and $R$. Given log-negativity of a pure state, it is possible to bound the mixed-state entanglement of a mixed state $\hat{\rho}$ (quantified by log-negativity of a mixed state $\mathcal{E}_{\text{N}}(\hat \rho)$) using the pure-state log-negativity and the notion of fidelity. In Ref.\,\cite{ShawNature2024}, it was shown that, if the desired pure state $|\psi\rangle$ is an eigenstate of the mixed-state density matrix $\hat \rho$ with an eigenvalue $F$, the following bound holds
\begin{gather}
    \label{eq:bound_logneg}
     \mathcal{E}_{\text{N}}(\hat \rho) \geqslant \mathcal{E}_{\text{N}}(|\psi\rangle) + \log_2 F,
\end{gather}
where $F = \langle \psi|\hat \rho|\psi\rangle$ is the state fidelity~\cite{ShawNature2024}. Notably, in an actual experiment, the pure state will not be an exact eigenstate of the density matrix. Nevertheless, Ref.\,\cite{ShawNature2024} has shown that the bound Eq.\,\eqref{eq:bound_logneg} holds for generic time evolution and generic noise sources. Log-negativity of a pure state $|\psi\rangle$ is equivalent to the Rényi--$1/2$ entropy and could be computed as 
\begin{gather}
    \mathcal{E}_{\text{N}}(\psi) = 2 \log_2 \sum\limits_{i = 0}^{2^L - 1} S_i,
\end{gather}
where $S_i$ are the Schmidt values.

In the $U(1)$--symmetric case, we obtain
\begin{gather}
    \sum\limits_{i = 0}^{2^L - 1} S_i \to \int\limits_{0}^{+\infty} \text{d}S\, S \,p(S) = \frac{D}{3\pi} \sum\limits_{M_L = 0}^{L} \lambda_+^{M_L} \times \\ 
    \nonumber \times
    \left[ \left[\left( \lambda_-^{M_L}\right)^2 + \left(\lambda_+^{M_L}\right)^2 \right] \tilde{E}(q) - 2 \left( \lambda_-^{M_L} \right)^2 \tilde{F}(q) \right],
\end{gather}
where $q = \sqrt{1 - \left(\lambda_-^{M_L} / \lambda_+^{M_L}\right)^2}$, and $\tilde{F}, \tilde{E}$ are complete elliptic integrals of the first and the second kind, respectively. In a system with $L = R = N / 2$ and even $M = (L + R) / 2$, the expressions simplify to $\lambda_-^{M_L} = 0$, $\lambda_+^{M_L} = 2 \sqrt{D_R^{M_L} / D}$ and $q = 1$ and the sum reads:
\begin{gather}
    \sum\limits_{i = 0}^{2^L - 1} S_i \to \frac{8}{3 \pi} \sum\limits_{M_L = 0}^{L} \frac{\left(D_R^{M_L}\right)^{3/2}}{D^{1/2}} = \\
    \nonumber = 
    \frac{2^{3/4} \times 8}{3 \sqrt{3} \pi} 2^{N/4} \left(1 + \mathcal{O}(1/N)\right),
\end{gather}
and we obtain
\begin{gather}
\label{eq:pure_log_neg}
    \mathcal{E}^{U(1)}_N = N/2 + \log_2 \frac{2^{3/2}}{3} \frac{64}{9 \pi^2} + \mathcal{O}(1/N) = \\
    \nonumber = 
    \frac{S_{\text{ent.}}^{U(1)}}{\log\, 2} + \frac{1}{2} + \frac{1}{4 \log 2} + \log_2 \frac{2^{3/2}}{3} \frac{64}{9 \pi^2} + \mathcal{O}(1/N) \approx \\ \nonumber \approx \frac{S_{\text{ent.}}^{U(1)}}{\log\, 2} + 0.303.
\end{gather}
In contrast, a system with the Hilbert space $D$ where the RDM does not form blocks, the log-negativity reads 
\begin{gather}
\label{eq:pure_log_neg_generic}
    \mathcal{E}_{\text{N}}^{\text{generic}} = \log_2 \sqrt{D} + \log_2 \frac{64}{9 \pi^2} = \\
    \nonumber = 
    \frac{S_{\text{ent.}}^{\text{generic}}}{\log 2} + \frac{1}{2 \log 2} + \log_2 \frac{64}{9 \pi^2} \approx \frac{S_{\text{ent.}}^{\text{generic}}}{\log 2} + 0.248.
\end{gather}
We use the expression for log-negativity in its $U(1)$--conserving (Eq.\,\eqref{eq:pure_log_neg}) and generic (Eq.\,\eqref{eq:pure_log_neg_generic}) forms in the main text to approximate the maximum mixed-state entanglement entropies obtained in different experiments. 

In the main text, Figure 2d, on the $x$-axis we plot the effective system size defined as 
\begin{gather}
    N^{\text{eff}}_{\text{q}} = 2 S_{\text{ent}} / \log 2 + 1 / \log 2,
\end{gather}
such that $N^{\text{eff}}_{\text{q}} = N_{\text{q}}$ for a generic system. In turn, in the $y$-axis, we plot $\mathcal{E}_{\text{N}}$ against $N^{\text{eff}}_{\text{q}} = 2 S_{\text{ent}} / \log 2 + 1 / \log 2$ and draw an ideal line. However, for the $U(1)$--conserving and generic cases, the correction to $\mathcal{E}_{\text{N}} = S_{\text{ent.}} / \log 2 + C$ is slightly different ($C_{U(1)} = 0.303$ against $C_{\text{generic}} = 0.248$, respectively). For simplicity, to plot the ideal line, we use the mean value $\bar{C}$ between these two corrections and plot $\mathcal{E}_{\text{N}} = (1/2) N^{\text{eff}}_{\text{q}} - 1 / (2 \log 2) + \bar{C} \approx (1/2) N^{\text{eff}}_{\text{q}} - 0.449.$

\begin{figure}[t!]
    \centering
    \includegraphics[width=1.0\columnwidth]{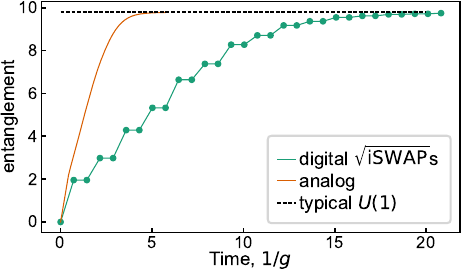}
    \caption{{\bf Comparison of entanglement growth in the digital and analog cases.} The points show entanglement entropy in the digital case after application of each layer in the ABCD gate pattern. The black dashed line shows the typical entanglement in the $U(1)$ case of a 30-qubit system obtained using Eq.\,\eqref{eq:Sent_U1}.}
    \label{fig:appendix_analog_vs_digital}
\end{figure}

\subsubsection{\texorpdfstring{Rényi-$2$ entropy}{Rényi-2 entropy}}
The Rényi-$2$ entropy could be experimentally obtained via randomized purity measurements~\cite{elben_2023}. 
Similarly to the Rényi-$1/2$ entropy, in the $U(1)$--symmetric case, entanglement of a typical quarter-circle state reads
\begin{gather}
    \sum\limits_{i = 0}^{2^L - 1} S_i^4 \to \int\limits_{0}^{+\infty} \text{d}S\, S^4 \,p(S) =  \\ \nonumber \frac{D}{32} \sum\limits_{M_L = 0}^{L} \left[\left(\lambda_+^{M_L}\right)^2 - \left(\lambda_-^{M_L}\right)^2 \right] \times \left[ \left(\lambda_+^{M_L}\right)^2 + \left(\lambda_-^{M_L}\right)^2 \right].
\end{gather}
Considering a system with $L = R$ and even $M = (L + R) / 2$, we obtain the $\log_2$--based expression
\begin{gather}
\label{eq:max_renyi_2}
    S^{U(1)}_{\text{Rényi-2 ent.}} = (N/2) + \log_2 \frac{\sqrt{3}}{4} + \mathcal{O}(1/N).
\end{gather}

\subsubsection{Entanglement growth in digital and analog cases}

In this subsection, we compare the rates of the bipartite von Neumann entropy growth in the digital and analog settings. We consider a two-dimensional $5 \times 6$ lattice with open boundary conditions and cut the lattice into two 15-qubit parts along the shorter direction. In the analog case, we simulate the pure $XY$-model with the coupling constant $g$. In the digital case, we consider a period-4 ABCD pattern (left, up, right, down) of $\sqrt{\text{iSWAP}}$ gates. Applying a layer of such gates requires time $t = (\pi / 4) (1/g).$ We can therefore put analog and digital simulations on the same time-axis. The result is shown in Fig.\,\ref{fig:appendix_analog_vs_digital}. We observe that both entropies reach the bound set by the quarter-circle theory, with the analog version being nearly 4 times as fast.

\subsection{\texorpdfstring{Rényi-$2$ entanglement entropy}{Rényi-2 entanglement entropy}}
\begin{figure}[h!]
    \centering
    \includegraphics[width=1.0\columnwidth]{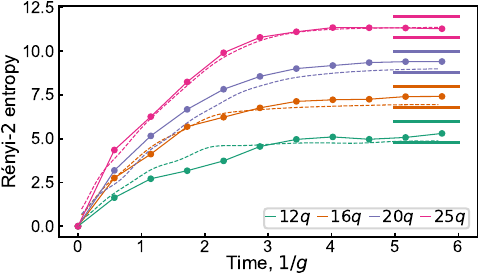}
    \caption{{\bf Measurement of the bipartite Rényi-$2$ entanglement entropy.} Solid lines are the experimental results, and the dashed lines are fidelity-adjusted simulation results. For each system size, the lower bold line on the right denotes entropy of a typical state with $U(1)$ conservation given by Eq.\,\eqref{eq:max_renyi_2}, and the upper bold line corresponds to the maximally-depolarized state at this system size.}
    \label{fig:SI_msmt_entanglement}
\end{figure} 

\subsubsection{Measurement}

In order to measure the entanglement entropy without fully reconstructing the density matrices, we utilize randomized measurement techniques~\cite{elben_2023}, and the results are shown in Fig.\,\ref{fig:SI_msmt_entanglement}. To probe various properties of the prepared quantum states, we apply $M$ sets of single-qubit Clifford gates prior to measurement to all qubits in our system, and repeat the measurement to obtain $K$ shots. Using this protocol we obtain $M \times K$ bitstrings, which we use to compute entanglement entropy. The state purity $P = \text{tr}(\hat \rho^2)$ is estimated using
\begin{equation}
    P = \frac{2^N}{M} \sum^{M}_{m=1} \sum^{2^N}_{s, s^\prime} P(s) P(s^\prime) (-2)^{-D[s, s^\prime]}\,,
\end{equation}
where $N$ corresponds to the number of qubits in the (sub)system, and $D[s, s^\prime]$ is the Hamming distance between the two bitstrings $s$ and $s^\prime$. In order to mitigate the bias in the estimation of purity arising from finite samples, we utilize jackknife resampling to obtain the unbiased purity estimate $\hat P_{\text{unbiased}}$ from the measured value $P$ using the following formula
\begin{equation}
    P_{\text{unbiased}} = \frac{K}{K - 1} \hat P - \frac{2^N}{K - 1}\,.
\end{equation}
Finally, we calculate the second Rényi entropy
\begin{equation}
    S_2(\hat \rho) = -\log_2 (P_{\text{unbiased}})
\end{equation}
to characterize the entanglement entropy of the (sub)system.

In order to probe the entanglement entropy of volume-law (typical) subsystems of up to size $12$ in our system, we use $M=50$ different pre-measurement unitaries, with $K=10^6$ shots each. For a typical state, the probability of measuring a given bitstring scales as $2^{-N}$. Therefore, the accurate reconstruction of the probability distribution requires many shots. The bitstring probability distribution does not vary by much when measured in different bases, hence, we only need to use a small $M$. In contrast, when probing area-law states we need larger $M$ whilst requiring fewer shots. In our experiments, we use $M=1000$ pre-measurement unitaries with $K=5 \times 10^4$ shots for such states.
In order to accurately probe the entanglement scaling crossover from area- to volume-law as a function of the number of excitations in the initial state $n_0$, we use $M=1000$ and $K=5 \times 10^4$ for states with $n_0 < 8$, and $M=50$ and $K=10^6$ for states with $n \geqslant 8$.

\subsubsection{Numerics}
To match the experimental data, we consider finite fidelity-corrections to the simulated Schmidt spectrum. To this end, we write the full system density matrix as 
\begin{gather}
    \hat \rho = F |\psi\rangle \langle \psi| + (1 - F) \frac{\text{Id}}{D},
\end{gather}
assuming the global depolarizing channel, where $\text{Id}$ is the identity matrix. If the system is split into the left and right parts with sizes $L$ and $R$, the RDM of the left-hand side reads
\begin{gather}
    \hat{\rho}_L = \bigotimes\limits_{M_L = 0}^{\text{min}(L, M)} 
    \left(F \hat{\rho}_{L, M} + \frac{1 - F}{D} D_R^{M_L} \text{Id} \right),
\end{gather}
where the direct product runs over the number of excitations in the left part, similarly to Eq.\,\eqref{eq:factorization}. Randomized purity measurements compute 
\begin{gather}
    \text{Tr}\,\hat{\rho}_L^2 = \sum\limits_{M_L = 0}^{\text{min}(L, M)} \text{Tr}\left[ F \hat{\rho}_{L, M} + \frac{1 - F}{D} D_R^{M_L} \text{Id} \right]^2,
\end{gather}
where $\hat{\rho}_{L, M}$ is the perfect system density matrix. This expression allows us to estimate the effect of finite fidelity on the Rényi entropy estimations, $-\log_2 \text{Tr}\,\hat{\rho}_L^2.$ In Fig.\,\ref{fig:SI_msmt_entanglement}, we show the corrected result, using the $F(N_{\text{q}}, tg)$ dependence obtained in Section\,\ref{appendix:fidelity_fitting}. 

\subsection{Classical computational complexity}
In this section, we address the classical simulation complexity of the full 69-qubit chip using tensor-network contractions and MPS simulations.

\subsubsection{Tensor network contraction}

In a tensor network contraction approach, a circuit is represented in the form of elementary tensors (gates) with legs that need to be contracted~\cite{Markov2008, Villalonga2019, Gray2021hyperoptimized}. In case of XEB benchmarking, we are interested in a particular amplitude $\langle s|\psi\rangle$, where $|\psi\rangle$ is the simulated wave function and $s$ is a bitstring sampled on a quantum device. If $\hat{U}(t)$ represents a unitary evolution performed on a quantum device starting from an initial state $|s_0\rangle$, we need to compute scalars of the form $\langle s|\hat{U}(t)|s_0\rangle$. While tensor contraction methods are directly applicable to the problems of random circuit sampling~\cite{boixo2018simulation, chen2018classical, pednault2020paretoefficient, huang2020classical, kalachev2022multitensor, Pan2022, Pan2022b}, to study analog evolution with tensor network contraction methods, we first find an efficient digital circuit that represents the time evolution.  For a fair comparison, the circuit should be chosen to minimize its contraction cost.

\begin{figure*}[t!]
    \centering
    \includegraphics[width=1.0\textwidth]{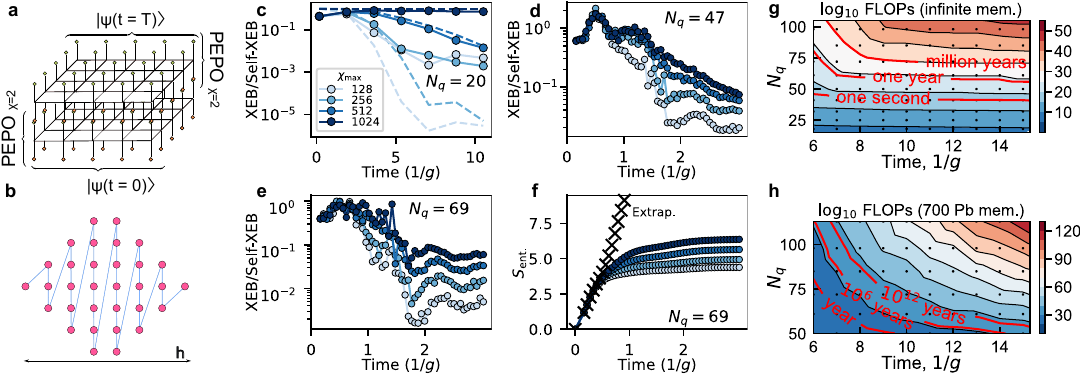}
    \caption{{\bf Classical computation complexity. a,} Representation of the hypothetical most efficient tensor network representation of the time evolution. Here, we show application of two PEPO layers to a wave function $|\psi(0)\rangle$ to obtain a wave function at time $t$, $|\psi(t)\rangle$, {\bf b,} A family of Sycamore-like geometries with varied height $h$ considered in this panel for complexity analysis. The blue line shows a typical MPS ``snake'' used to build an MPS wave function. {\bf c, }  The ratio of XEB for $N_{\text{q}}=20$, computed from experimental data and MPS simulations, to self-XEB, computed by sampling bitstrings from the MPS. Dashed lines show the fidelity of MPS wave functions relative to the exact time evolution. When the classical results are exact ($\chi=1024$), XEB/Self-XEB is a measure of experimental fidelity; at smaller $\chi$, we find a lower ratio. The $x$-axis gives the time in inverse coupling strength, taking into account weaker coupling during ramps. {\bf d,} XEB/Self-XEB for $N_{\text{q}}=47$. MPS fidelity at $\chi=1024$ is high until time approximately $1/g$, then decays rapidly.
    Self-XEB at time $3/g$ extrapolates in $1/\chi$ to around 2, so that the PTD would not yet be achieved.
    {\bf e,} The $N_{\text{q}}=69$ system. {\bf f,} Growth of entanglement in the MPS simulation of the 69-qubit system.  At short times, an extrapolation in $1/\log(\chi)$ gives the expected linear growth of entanglement. {\bf g, } Computational complexity of the TN-contraction algorithm, assuming no memory constraints, and {\bf h,} assuming using the full Frontier hard drive of 700 PB~\cite{frontier}. The red lines convert the FLOPs count in Frontier time assuming its peak 2 exaFLOPs performance~\cite{frontier}.}
    \label{fig:SI_complexity}
\end{figure*}

We assume that the gates in a digital circuit representing the time evolution can be collected, layer-by-layer, into projected entangled-pair operators (PEPOs). The resulting tensor network is shown in Fig.\,\ref{fig:SI_complexity}a. For concreteness, we specifically assume that the time evolution can be written in terms of PEPOs with virtual bond dimension $\chi = 2$, and that these PEPOs are maximally efficient in terms of generating entanglement. Given the virtual bond dimension, a single PEPO application generates $\Delta S_{\text{ent.}} = L_{\text{short}} \log\,2$ entanglement entropy across the minimal cut that divides the system in half. Since the maximum entanglement entropy is $S_{\text{ent.}}^{\text{max}} = (1/2) L_{\text{short}} L_{\text{long}} \log\,2$, it takes at least $N_{\text{PEPOs}} = (1/2) L_{\text{long}}$ layers of PEPO application to saturate the entanglement entropy. Note that considering larger bond dimension of the form $\chi = 2^k$ does not change this consideration. Indeed, it would generate the same amount of entanglement as $k$ PEPOs with $\chi = 2$, and also could be written as consecutive application of $k$ such PEPOs.

As shown in Sec.\,\ref{subsec:entanglement}, the entanglement entropy saturates in $(1/2 g) \times L_{\text{long}}$. Therefore, a single PEPO application generates the amount of entropy which corresponds to the evolution time of a single cycle ($1 /g$) defined in the main text. This conversion allows us to estimate the contraction costs, considering a set of Sycamore-like tilted square geometries shown in Fig.\,\ref{fig:SI_complexity}b. 

In Fig.\,\ref{fig:SI_complexity}g, we show the FLOPs count required to contract a tensor network with $N_{\text{q}}$ qubits and a given number of analog cycles, assuming infinite memory. Here, we do not account for finite fidelity of the device. The best contraction is found using a simulated annealing algorithm repeated with numerous attempts.

Finally, in Fig.\,\ref{fig:SI_complexity}h, we show complexity assuming usage of the whole Frontier hard drive (700 PB)~\cite{frontier} and ignoring communication costs. Finite memory limits the size of a maximum tensor that could be stored in memory during the computation. To avoid having a larger tensor, some edges in the tensor network are projected and then the final result is summed over all $2^{N_{\text{sliced}}}$ possible choices of $N_{\text{sliced}}$ sliced variables~\cite{chen2018classical, Villalonga2019, Huang2021slicing}. To account for finite fidelity of the simulation, we only consider the respective fraction of the sliced variables' combinations~\cite{Villalonga2019}. We observe that contraction of a circuit representing evolution on a 69-qubit chip until the maximum-entanglement time would require $\mathcal{O}(10^6)$ years on the Frontier supercomputer assuming its peak 2 exaFLOPs performance~\cite{frontier}.

\subsubsection{Matrix product states---complexity bounds}
In this subsection, we estimate the complexity of sampling bitstring form the distribution $|\langle s|\hat{U}(t)|s_0\rangle|^2$, using matrix product states (MPS). We consider the ``MPS snake'' site encoding, as shown in Fig.~\ref{fig:SI_complexity}b. In this section, the entanglement and corresponding MPS bond dimension, $\chi$, are considered for a half-system cut. In addition to the experimental fidelity $\mathcal{F}_{\text{exp}}$, we consider fidelity $\mathcal{F}_{\chi_{\text{max}}}$ of the time-evolved MPS with the peak bond dimension at most $\chi_{\text{max}}$. Given the total time-evolution length $t$, we need to estimate the minimum peak bond dimension $\chi_{\text{max}}$ such that (i): $\mathcal{F}_{\chi_{\text{max}}}(t) \geqslant \mathcal{F}_{\text{exp}}(t)$. This requirement only at time $t$ is weaker than demanding (ii): $\mathcal{F}_{\chi_{\text{max}}}(\tau) \geqslant \mathcal{F}_{\text{exp}}(\tau)$ for all $0 \leqslant \tau \leqslant t$, and therefore (i) gives a lower bound on (ii).

We choose the total evolution time $t$ such that satisfying criterion (i) with MPS is as hard as possible. In the experiment, entanglement entropy builds up linearly and saturates at $t_{\text{sat.}} \sim L_{\text{long}} / (2 g)$, as shown in Fig.\,\ref{fig:SI_entanglement}. At $t > t_{\text{sat.}}$, $F_{\text{exp.}}(t)$ decays exponentially, which makes $t = t_{\text{sat.}}$ a natural choice. However, a single bitstring amplitude $\langle s|\hat{U}(t)|s_0\rangle$ can be obtained by first obtaining $\hat{U}(t/2)|s_0\rangle$ and $\hat{U}(-t/2)|s\rangle$, and then finding their overlap, so we may only time-evolve MPS to half-time. Therefore, we choose $t = 2 t_{\text{sat.}}.$ On a 69-qubit system with $L_{\text{long}} = 8$, $2 t_{\text{sat.}} \sim 8 / g$. As shown in Section.\,\ref{appendix:fidelity_fitting}, the extrapolated experimental fidelity at this time is $\mathcal{F}_{\text{exp}} (2 t_{\text{sat.}}) \approx 0.4$. To achieve such fidelity at $t / 2 = t_{\text{sat.}},$ an MPS would employ the bond dimension of $\chi_{\text{max}}\approx 1.7\times 10^9$. This would require around 50$\times$ the total hard drive of Frontier for the largest individual tensors. This estimate did not account for the $U(1)$ conservation.

With the $U(1)$ symmetry, the MPS tensors have a block structure~\cite{Singh2011}. We find numerically for the 69 qubit system that at bond dimensions of $2^7$, $2^8$, $2^9$, and $2^{10}$, at $t=3 / g$ the required memory is multiplied by $\approx$ 0.20, 0.19, 0.18, and 0.17, respectively. A linear extrapolation in $1/\log(\chi)$ shows that the memory requirement for large bond dimensions is reduced by $m_{U(1)} \sim 10$, therefore the largest MPS tensors would still exceed the Frontier hard drive. 

Finally, let us assume no memory constraint and estimate the pure FLOPs requirement to perform such MPS simulation, which is dominated by the singular value decomposition (SVD).
For a real $n\times n$ matrix, a SVD takes $\mathcal{O}(n^3)$ operations, and the MPS compressions involve truncation of $2\chi\times 2\chi$ complex matrices due to inclusion of the physical dimension of 2 on each site, leading to the $64\chi^3$ FLOPs requirement per SVD. The $U(1)$ symmetry reduces the cost by a factor $m_{U(1)}^2$. We will conservatively only include the cost of SVD at the central cut with largest bond dimension, ignoring decompositions at other cuts.

Assuming a single cycle requires $N_{\text{Trotter}}$ steps, we get the total number of SVD decompositions $N_{\text{SVD}} = 4 N_{\text{Trotter}}$. We assume that the entanglement increases linearly during the simulation up to time $t_{\text{sat.}}$, and therefore the required bond dimension grows at step $k$ as $\chi(k) \sim (\chi_{\text{max}})^{k / N_{\text{SVD}}}$. The required FLOPs estimate reads
\begin{gather}
    \text{FLOPs} = \frac{64}{m_{U(1)}^2} \sum\limits_{k = 0}^{N_{\text{SVD}}} \left((\chi_{\text{max}})^{k / N_{\text{SVD}}}\right)^3 \sim \\ \nonumber \sim \frac{64}{m_{U(1)}^2} \frac{N_{\text{SVD}}}{3 \log \chi_{\text{max}}} \chi_{\text{max}}^3.
\end{gather}

Maintaining low Trotter error requires at least 5 sweeps per cycle, which gives $\text{FLOPs} \sim 10^{27}$ (with $\chi_{\text{max}}\approx 1.7\times 10^9$), which would take $0.5 \times 10^9\,\text{s} = 16\,\text{years}$ on Frontier assuming its peak performance of 2 exaFLOPs and ignoring any communication costs.

Importantly, in this procedure with forward and backward evolution, final bitstrings $s$ will be chosen from i.i.d. and will not follow the PTD. Rejection sampling\,\cite{markov2018quantum} corrects for this at the cost of an extra $\mathcal{O}(10)$ sampling overhead. Thus the effective time per bitstring for MPS is lower-bounded by $160$ years.

\subsubsection{Matrix product states --- practical demonstration}
We run MPS time evolution simulations for the system sizes $N_{\text{q}}=20,$ 47, and 69; we show the results in Fig.~\ref{fig:SI_complexity}c-f\footnote{Similar results were found for $N_{\text{q}}=30$ and 60.}. The MPS sites are ordered as shown in Fig.~\ref{fig:SI_complexity}b. We account for the $U(1)$ conservation and construct a near-optimal matrix product operator (MPO) using deparallelization and delinearization~\cite{Hubig2017}. We time-evolve using the two-site time-dependent variational principle (TDVP)~\cite{Haegeman2016, PAECKEL2019}, using time steps $\delta t \leqslant 0.05/g$ so that time discretization is not a significant source of error. We simulate the full time-dependent experimental procedure, including ramps and plateau time.
For each system size, we consider bond dimensions $\chi\in\{128, 256, 512, 1024\}$, with truncation after each local TDVP step. To track the MPS fidelity, we either (i) record truncation error at each step, or (ii) compute the fidelities between the same-time states with the various bond dimensions.

For $N_{\text{q}}=20$, where $\chi = 1024$ yields no truncation, and (ii) is exact, the results are shown as dashed lines in Fig.~\ref{fig:SI_complexity}c. Similarly, for all considered system sizes, MPS simulations at our largest bond dimensions maintain high fidelity until times $\sim 1/g$, followed by an exponential decay with the rate higher than the experimental (see Sec.~\ref{appendix:fidelity_fitting}).

We consider the ratio of XEB to self-XEB, gives an estimate of fidelity (see Sec.~\ref{sec:XEB_benchmarking_fidelity}). To obtain linear XEB, we perform MPS evolution until time $t$ with bond dimension $\chi$, and compute 
\begin{gather}
    \text{XEB}~=~\frac{D}{M} \sum\limits_{x \sim p_{\rm meas}(x,t)} p_{\rm MPS}(x,t) - 1    
\end{gather}
with $x$ being the experimental bitstrings, and 
\begin{gather}
    \text{self-XEB}~=~\frac{D}{M} \sum\limits_{x \sim p_{\rm MPS}} p_{\rm MPS}(x, t)-1,
\end{gather}
where $x$ are sampled from an MPS~\cite{Stoudenmire_2010, Ferris2012}, and $M$ is the total number of samples.

At $N_{\text{q}}=20$, experimental fidelity is above $0.5$ for times up to $10 / g$, and the MPS simulations at $\chi=1024$ are exact. Indeed, in Fig.~\ref{fig:SI_complexity}c, we observe XEB/self-XEB being close to 1 for $\chi=1024$. Furthermore, for $\chi=512$, XEB/self-XEB is close to the MPS fidelity; since the experimental fidelity is high, this is consistent with XEB/self-XEB giving an approximate relative fidelity between the experiment and the MPS. When the MPS fidelity is lower, the XEB/self-XEB ratio provides no clear information, as plots with $\chi = 128$ and $256$ show.

With these insights, we can use MPS to examine XEB/self-XEB for systems of 47 and 69 qubits. Generally, rapidly-growing entropy does not allow for reliable fidelity decays extrapolations. However, for $N_{\text{q}}=47, 69$, shown in Figs.~\ref{fig:SI_complexity}d-e, the MPS fidelity for the largest bond dimension remains close to 1 until plateau time $\approx 1/g$. In this regime the MPS is not significantly truncated (there is a clear convergence towards larger bond dimensions), and thus the ratio XEB/self-XEB gives an approximate handle on the experimental fidelity (at these times, the self-XEB is still of order 100, so the distribution of bitstrings is far from PTD, making the connection only approximate). Therefore, $\text{XEB/self-XEB} \sim 1$ in Fig.~\ref{fig:SI_complexity}d-e verifies that the quantum device is behaving similarly to the simulation, and that the experimental fidelity remains high at least within the first cycle.
At longer times, although the MPS fidelity decays rapidly, the XEB/self-XEB is converging with increasing bond dimension. In principle, an extrapolation in bond dimension could allow a quantitative estimate of fidelity decay rate at longer times.

In addition to fidelity estimates, at short times the entanglement entropy between two halves of the system can be computed.  In Fig.~\ref{fig:SI_complexity}f we show the entanglement vs time at different bond dimensions for the full 69-qubit system.  We find the expected near-linear growth in entanglement at short times before reaching the plateau.  We linearly extrapolate in $1/\log(\chi)$ to infinite bond dimension, showing that linear growth of entanglement would continue as expected if we used larger $\chi$. The short-time entanglement entropies for $\chi=1024$ with $N_{\text{q}}=47$ and 69 are also shown in Fig.~\ref{fig:SI_entanglement}e.

\begin{figure*}[t!]
    \centering
    \includegraphics[width=1.0\textwidth]{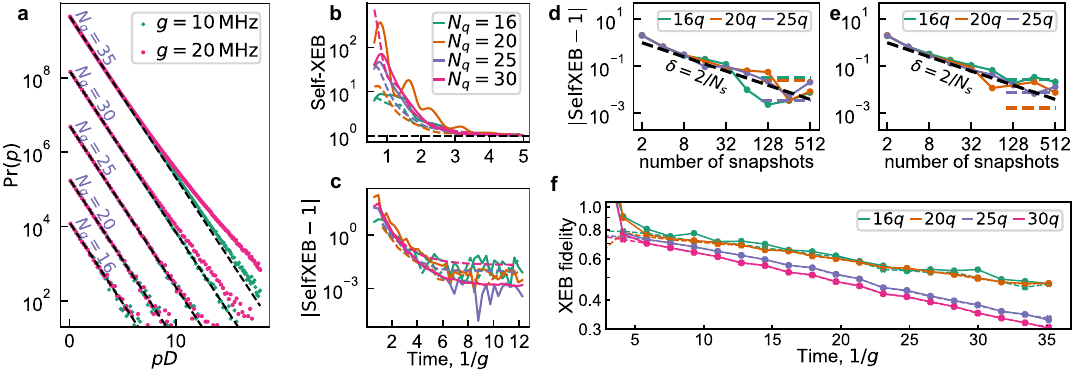}
    \caption{{\bf Bitstring distributions of simulated wave functions. a,} The probabilities distribution of the simulated wave functions for $g / (2 \pi) \approx 10\,\text{MHz}$ and $g / (2 \pi) \approx 20\,\text{MHz}$. The black dashed lines correspond to the ideal Porter-Thomas distribution. {\bf b,} Convergence of self-XEB in analog evolution to $1$ for $g / (2 \pi) \approx 10\,\text{MHz}$ (bold) and $g / (2 \pi) \approx 20\,\text{MHz}$ (dashed). {\bf c,} Time-dependence of the imperfection $\delta = |\text{self-XEB} - 1|$. 
    {\bf d, e,} The imperfection after $p_{\text{avg.}}(s_i)$-renormalization as a function of the number of snapshots $N_s$ for 
    $g / (2 \pi) \approx 10\,\text{MHz}$ and $g \approx 20\,\text{MHz}$, respectively. The black dashed line shows the theoretical prediction $\delta = 2 / N_s + \mathcal{O}(1/N_s^2),$ while the colored dashed line show the imperfections for different system sizes without the renormalization. {\bf f,} Linear XEB fidelity as a function of time measured in $1/g$ for various system sizes for $g / (2 \pi) \approx 10\,\text{MHz}$, the dataset used in the main text. The solid lines show XEB estimator $F$ before renormalization, while dashed lines show the XEB estimator $\tilde F$ after renormalization.}
    \label{fig:SI_distributions}
\end{figure*}

\subsection{Bitstring distributions and XEB}
In this section, we provide evidence that the simulated wave functions to a large degree follow the Porter-Thomas distribution (PTD) $\text{Pr}(p) = D e^{-p D}$, where $D$ is the system Hilbert space dimension. The PTD is a key element in benchmarking the state fidelity using XEB~\cite{morvan2023phase}. We will also study the applicability of the recently suggested protocols to remove the non-PTD corrections from the distribution of bitstrings in a wave function obtained in analog evolution.

\subsubsection{Distributions and self-XEB}
We give the summary of results in Fig.\,\ref{fig:SI_distributions}. In Fig.\,\ref{fig:SI_distributions}a, we plot the bitstring probability distributions of the wave functions obtained in half-filled simulation of the $g / (2 \pi) \approx 10\,\text{MHz}$ and $g / (2 \pi) \approx 20\,\text{MHz}$ cases for system sizes varying between 16 and 35 qubits. The wave functions are taken, respectively, after 300 and 150\,ns of time evolution, corresponding to 18 cycles. The Hilbert space dimension $D = {N_{\text{q}} \choose N_{\text{q}} / 2}$ accounts for the particle number conservation. As the system size increases, we observe slight deviation of the bitstring distributions from the PTD, which becomes worse in the $g / (2 \pi) \approx 20\,\text{MHz}$ case, where non-XY terms have greater magnitude. The degree of agreement with the PTD could be quantified by self-XEB $D \sum_s p^2(s) - 1$, where $p^2(s)$ is the simulated probability of a bitstring $s$, and the sum runs over the whole Hilbert space. In case of a PTD, self-XEB is unity. Generally, self-XEB could vary from $0$ for a fully-depolarized state, to $D - 1$ for a fully-localized state. In Fig.\,\ref{fig:SI_distributions}b, we show self-XEB as a function of time for different system sizes and $g / (2 \pi) \approx 10,\,20\,\text{MHz}$. It reaches $\mathcal{O}(1)$ after a short time of $t \leqslant 4/g$. In Fig.\,\ref{fig:SI_distributions}c, we show the absolute deviation of self-XEB from 1, $\delta$ as a function of time. We observe that, as seen in Fig.\,\ref{fig:SI_distributions}a, the $g / (2 \pi) \approx 20\,\text{MHz}$ simulation has a larger deviation (up to $\delta \sim 0.03$) from the PTD, as compared to the $g / (2 \pi) \approx 10\,\text{MHz}$ case, where PTD is nearly-reached.

\subsubsection{XEB benchmarking of fidelity}
\label{sec:XEB_benchmarking_fidelity}
In this section, we will discuss benchmarking of the state fidelity using XEB. To this end, we use a simple model for the system density matrix
\begin{gather}
    \hat{\rho} = \Phi |\psi_{\text{exp.}}\rangle \langle \psi_{\text{exp.}}| + (1 - \Phi) \frac{\text{Id}}{D},
\end{gather}
where $\text{Id}$ is the identity matrix representing the global depolarizing noise channel, $|\psi_{\text{exp.}}\rangle$ is the true wave function that describes the device evolution, and $\Phi$ represents fidelity of the time evolution with respect to non-coherent errors. Additionally, due to the calibration imperfections, the simulated wave function $|\psi_{\text{sim.}}\rangle$ is different from $|\psi_{\text{exp.}}\rangle$, which is realized on the chip. We could express the simulated wave function as
\begin{gather}
    |\psi_{\text{sim.}}\rangle = \sqrt{f} |\psi_{\text{exp.}}\rangle + \sqrt{1 - f} |\psi_{\perp}\rangle,
\end{gather}
where $\langle \psi_{\text{exp.}} |\psi_{\perp}\rangle = 0$ and $f = |\langle \psi_{\text{exp.}} | \psi_{\text{sim.}} \rangle|^2$ represents fidelity with respect to coherent errors. With these definitions, the experimental probabilities read
\begin{gather}
    p_{\text{exp.}}(s) = \langle s|\hat{\rho}|s\rangle = p_s \Phi + (1-\Phi) / D
\end{gather}
with $p_s = |\psi_{\text{exp.}}(s)|^2$, and the simulated probabilities read
\begin{gather}
    p_{\text{sim.}}(s) = |\psi_{\text{sim.}}(s)|^2 = p_s f + \\ \nonumber + 2 \text{Re}\left[\sqrt{f (1 - f)} \psi_{\text{exp.}}(s) \psi^*_{\perp}(s)\right] + (1 - f) p_{\perp, s}
\end{gather}
with $p_{\perp, s} = |\psi_{\perp}(s)|^2$. The XEB then reads
\begin{widetext}
    \begin{gather}
    \label{eq:XEB_bare}
        \text{XEB} = D \sum\limits_{s} p_{\text{exp.}}(s) p_{\text{sim.}}(s) - 1 =  D \sum_s \left[f \Phi p^2_s + \Phi p_s \left( 2 \text{Re}\left[\sqrt{f (1 - f)} \psi_{\text{exp.}}(s) \psi^*_{\perp}(s)\right] + (1 - f) p_{\perp, s}\right) \right. + \\ \nonumber + \left[\frac{1 - \Phi}{D} \left(p_s f + 2 \text{Re}\left[\sqrt{f (1 - f)} \psi_{\text{exp.}}(s) \psi^*_{\perp}(s)\right] + (1 - f) p_{\perp, s} \right) \right].
    \end{gather}
\end{widetext}

We work out these terms and then combine them into the final expression. First, we note the simple relations such as
\begin{gather}
    D \sum\limits_s p_s^2 = \text{self-XEB} + 1,\;
    \sum\limits_s p_{\perp, s} = \sum\limits_s p_s = 1, \\
    D \sum\limits_s p_s p_{\perp, s} = D^2 \mathop{\mathbb{E}} p_s \times \mathop{\mathbb{E}} p_{\perp, s} = 1,
\end{gather}
where we treated $p_s$ and $p_{\perp, s}$ as {\it uncorrelated} random variables. In order to simplify the cross-terms, we write $\text{Re}\,\left[\psi_{\text{exp.}}(s) \psi_{\perp}^*(s)\right] = \sqrt{p_s p_{\perp, s}} \cos (\text{arg}\, \psi_{\text{exp.}}(s) - \text{arg}\, \psi_{\perp}(s))$. Assuming both probability distributions $p_s$ and $p_{\perp, s}$ nearly follow the PTD, and treating $\delta \theta(s) = \text{arg}\, \psi_{\text{exp.}}(s) - \text{arg}\, \psi_{\perp}(s)$ as an uncorrelated random variable, we obtain 
\begin{gather}
    \mathop{\mathbb{E}} \sqrt{p_s p_{\perp, s}} \cos \theta_s = 0,\; \mathop{\mathbb{E}} \left(\sqrt{p_s p_{\perp, s}} \cos \theta_s\right)^2 = 1 / (2 D^2),
\end{gather}
similarly 
\begin{gather}
    \mathop{\mathbb{E}} \sqrt{p_s^3 p_{\perp, s}} \cos \theta_s = 0,\; \mathop{\mathbb{E}} \left(\sqrt{p_s^3 p_{\perp, s}} \cos \theta_s\right)^2 = 3 / D^4.
\end{gather}

Under these assumptions, the full XEB expression simplifies to 
\begin{gather}
    \text{XEB} = f \Phi (\text{self-XEB} + 1) + \Phi (1 - f) + (1 - \Phi) f + \\ \nonumber + (1 - f)(1 - \Phi) = f \Phi \text{self-XEB} = F \times \text{self-XEB}, 
\end{gather}
where we defined $F = f \Phi$, fidelity reflecting coherent and non-coherent errors. Therefore, we obtain 
\begin{gather}
    \label{eq:XEB_F}
    \text{XEB} / \text{self-XEB} = F,
\end{gather}
which coincides with the estimator $F_e$ from Ref.\,\cite{BenchmarkingSoonwonTheoryPRL2023}.

\subsubsection{Correlated probabilities in analog dynamics}
Above, we treated $p_s$ and $p_{\perp, s}$ as uncorrelated variables. However, the probability distributions $p_s$ and $p_{\perp, s}$ obtained from analog dynamics with conservation laws may retain a large degree of correlation, which would introduce corrections to the relation Eq.\,\eqref{eq:XEB_F}, since $\mathop{\mathbb{E}} p_s p_{\perp, s} \neq \mathop{\mathbb{E}} p_s \mathop{\mathbb{E}} p_{\perp, s}$. A symptom of these residual correlations is that the $p_s$ distribution has self-XEB $\neq 1$, i.\,e., imperfect agreement with the PTD. 

The information about the underlying analog dynamics and conservation laws is contained in the time-averaged probability distribution
\begin{gather}
    p_{\text{avg.}}(s) = \lim\limits_{T \to \infty} \frac{1}{T} \int\limits_0^{T} dt\, |\psi_{\text{sim.}}(t, s)|^2,
\end{gather}
where $|\psi_{\text{sim.}}(t, s)|^2$ is the $s$-bitstring probability for the simulated wave function at time $t$. 
Having obtained $p_{\text{avg.}}(s)$ by averaging over $N_s$ wave function snapshots, we could renormalize any probability distribution as $p_s \to p_s / (p_{\text{avg.}}(s) D)$, and re-weight the expectation values as $\mathop{\mathbb{E}} X_s = \sum\limits_s p_{\text{avg.}}(s) X_s$, thus removing the prevalence of certain bitstrings due to the details of the particular Hamiltonian dynamics implemented by the analog device~\cite{BenchmarkingSoonwonTheoryPRL2023}. 

In the limit of large sample $N_s$, the renormalized self-XEB $ = D \sum_s p_{\text{avg.}}(s) (p_s / p_{\text{avg.}}(s))^2 - 1$ converges to $1$, and the non-universal deviations from the PTD are removed. However, estimation of $p_{\text{avg.}}$ with finite $N_s$ leads to sampling noise, which for insufficient $N_s$ could make the imperfection $\delta = |\text{self-XEB} - 1|$ actually larger than if we did not perform the re-weighting at all.  To estimate the required sample size in order to achieve a given error level $\delta$, we consider a model case where the renormalization is applied to the PTD. We assume that $p_{\text{avg.}}(s)$ is estimated over $N_s$ samples sufficiently separated in time, such that all $p_{t, s} = |\psi_{\text{sim.}}(t, s)|^2$ could be seen as uncorrelated random variables drawn from PTD: $p_{t, s} \sim D e^{-p_{t, s} D}$. The inverse mean $1 / p_{\text{avg.}}$ of these $N_s$ samples is distributed as 
\begin{gather}
    1/p_{\text{avg.}} \sim \frac{N_s D p_{\text{avg.}}^2 (N_s D p_{\text{avg.}} )^{N_s - 1}} {(N_s - 1)!} e^{-N_s D p_{\text{avg.}} },
\end{gather}
giving $\langle 1/ [D p_{\text{avg.}}(s)] \rangle = 1 + 1/N_s + \mathcal{O}(1/N_s^2).$ Therefore, $\text{self-XEB} = 1 + 2/N_s + \mathcal{O}(1/N_s^2)$ for the case of the PTD renormalized with a finite sample $N_s$. As a result, for a non-PT distribution with $|\text{self-XEB} - 1| = \delta$ before re-weighting, $|\text{self-XEB} - 1|$ will be reduced by the re-weighting procedure if $N_s \leqslant 2/\delta$. We verify this scaling in Fig.\,\ref{fig:SI_distributions}d,e, considering renormalization of the analog distributions using a range of sample sizes $N_s$. Since the distributions produced in our analog dynamics are already close to the PTD, $\delta$ follows a $2/N_s$ trend robustly for both $g  / (2 \pi)\approx 10,\,20$\,MHz. The crossing with the non-renormalized value of $\delta$ happens at $N_s \sim 2 / \delta$, and the self-XEB improves afterwards. 

Having considered improvement of the analog distributions in terms of self-XEB, we now consider application of this  technique to the fidelity estimator $F = \text{XEB} / \text{self-XEB}$. If both experimental and theoretical distributions follow PTD, the renormalized version reads
\begin{gather}
    \label{eq:Fe}
    \tilde F = \frac{\sum_s p_{\text{exp}}(s) p_{\text{sim}}(s) / p_{\text{avg}}(s) - 1}{\text{self-XEB}} = \\ \nonumber =  F \left(1 - \frac{1}{N_s}\right) + \frac{1}{N_s} + \mathcal{O}(N_s^{-2}),
\end{gather}
which means that the finite sample size $N_s$ would lead to overestimation of fidelity. Therefore, considering simulated $|\text{self-XEB} - 1| \sim 0.01$, we apply this renormalization technique to the main dataset of the main text, averaging over 500 wave function instances to obtain $p_{\text{avg.}}(s)$. After renormalization, we compute the renormalized fidelity $\tilde F$, and the results are shown in Fig.\,\ref{fig:SI_distributions}f. We observe that the naive estimator $F$ (solid) agrees with the renormalized estimator $\tilde F$ (dashed). Therefore, we conclude that removing Hamiltonian-specific bias by means of $p_{\text{avg.}}(s)$ renormalization leaves the XEB decay rate unchanged, which demonstrates that our XEB proxy for fidelity is not affected by significant correlations between the probabilities $p_s$ and $p_{\perp, s}$.

\begin{figure}[t!]
   \centering
   \includegraphics[width=\columnwidth]{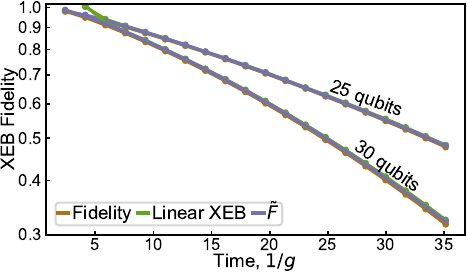}
   \caption{{\bf Representation of fidelity by XEB fidelity estimators} The curves represent fidelity, linear XEB and the $\tilde F$ estimator between the ``experimental'' and ``simulated'' states, which differ by coherent errors in the Hamiltonian. As a reference state, we pick 25- and 30-qubit wavefunctions considered in the main text. To compute $p_{\text{avg.}}$, we average over 501 wave function snapshots from 100 to 600 ns with a step of 1 ns.}
   \label{fig:appendix_F_representation}
\end{figure}

\subsubsection{Agreement between fidelity and estimators}
In this section, we show that the fidelity estimators used in this work faithfully represent the state fidelity. Here, we only consider the fidelity loss due to coherent errors, i.\,e., caused by calibration or modelling imperfections, which are shown to dominate the error in Sec.~\ref{appendix:selfXEB} . To this end, we consider (i) the wave functions $\psi_{\text{opt.}}$ at the optimal global shifts $\mathrm{d}g^{nXX} = \mathrm{d}g^{XnX} = -20\,\text{kHz}$ used in the main text at $g / (2 \pi) \approx 10\,\text{MHz}$ to benchmark experimental fidelity, and (ii) the wave functions $\psi_{\text{bare.}}$ without these shifts applied. We treat the $\psi_{\text{opt.}}$ wave functions as the ground truth $\psi_{\text{exp.}}$, and $\psi_{\text{bare.}}$ as our simulation $\psi_{\text{sim.}}$.

This allows us to evaluate how well linear XEB and the renormalized estimator, $\tilde F$ (Eq.\,\eqref{eq:Fe}) represent fidelity $|\langle \psi_{\text{exp.}} | \psi_{\text{sim.}}\rangle|^2$. The results are shown in Fig.\,\ref{fig:appendix_F_representation}. In agreement with our previous findings of the near-PTD shape of our probability distributions, we obtain good representation of fidelity by both naive linear XEB and its renormalized version Eq.\,\eqref{eq:Fe}.

\subsection{MPS simulations of Kibble-Zurek and diffusion experiments}
\begin{figure}
   \centering
   \includegraphics[width=\columnwidth]{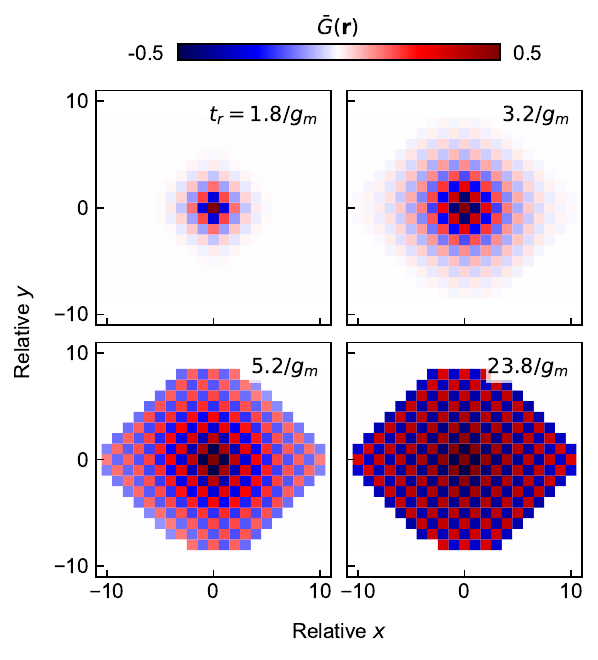}
   \caption{{\bf Averaged correlations at different ramp times.} Simulated equivalent of Fig.~3c of the main text showing the increase of the average correlations $\bar{G}(\mathbf{r})$ with increasing ramp time.}
   \label{fig:mps_space_corrs}
\end{figure}
To classically compute the time evolution of the state for the Kibble-Zurek experiments with $N_{\text{q}}=65$ sites, we perform time-dependent matrix product state simulations.  
Concretely, we employ the time-dependent variational principle (TDVP)~\cite{Haegeman2016} with two-site update to capture the buildup of bond dimension necessary when starting from the initial product state. We use the full Hamiltonian describing the experimental setup including next-nearest neighbor couplings and three-qubit terms illustrated in Fig.~\ref{fig:SI3}, and order the sites along the 1D MPS snake as depicted in Fig.~\ref{fig:SI_complexity}b. As a time step in the simulations, we choose $t_{\text{r}}/1000$, which translates to a maximal time step of $0.0625$ in units of $1/g_{\mathrm{m}}$ for the longest ramp times.
\begin{figure}
   \centering
   \includegraphics[width=\columnwidth]{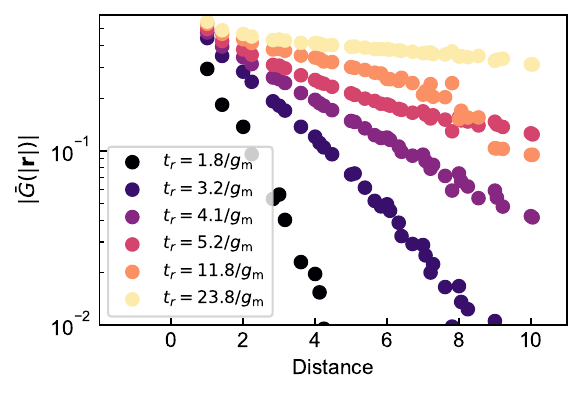}
   \caption{{\bf Averaged correlations as a function of distance at different ramp times.} The correlations at different distances agree well with the experimental data shown in Fig.~3d. As in the experiment, there is a crossover between exponential decay at small ramp times and power law behavior at large $t_{\text{r}}$.}
   \label{fig:mps_dist_corrs}
\end{figure}

\begin{figure}
   \centering
   \includegraphics[width=\columnwidth]{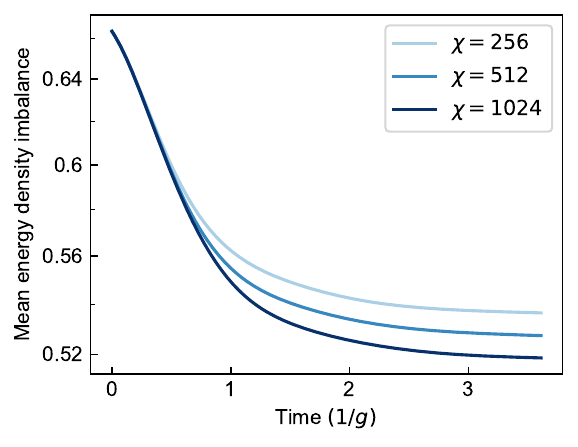}
   \caption{{\bf Breakdown of MPS simulations for energy diffusion.} The energy imbalance between the halves of the system prepared in high and low energy dimer coverings, respectively, first falls off exponentially as expected by diffusion (axis in log scale), but then plateaus due to the lack of bond dimension to faithfully represent the time-evolved state. The simulation already fails to capture the energy evolution at a time of $\sim 1/g$ at the highest bond dimension of $\chi=1024$.}
   \label{fig:mps_diff}
\end{figure}

This, on the one hand, ensures the absence of any significant error due to the finite time step, and, on the other hand, gives identical discrete increments of the time-dependent Hamiltonian along the ramp for all $t_{\text{r}}$. We simulate the dynamics for bond dimensions of up to $\chi=1024$ which leads to sufficient convergence of the observables measured in the experiment. The ability to obtain accurate results using MPS simulations for this large system size benefits from the interplay of two factors. If the ramp time is short, the system has only limited time to generate entanglement. If the ramp time is longer, the time-evolved state remains close to the ground state, guaranteeing area-law behavior with a logarithmic correction~\cite{metlitski2015entanglement}. The growth of entanglement is limited in both cases, making the MPS approach ideal to simulate this set of experiments. 

As already presented in the main text for a number of observables, this leads to a good agreement between experiment and simulation. Here, we present some additional numerical data. In Fig.~\ref{fig:mps_space_corrs}, we show the simulated version of Fig.~3c of the main text which agrees well with the experimental data.
The behavior of $\bar{G}(\mathbf{r})$ as a function of distance is depicted in Fig.~\ref{fig:mps_dist_corrs}. We observe very accurate agreement with experiment for smaller ramp times. For larger ramp times, the longer-range correlations in the experiment are slightly suppressed due to decoherence, while they become almost constant in the simulation.

We also attempted to simulate the energy transport experiment in which two halves of the system are prepared with high and low energy dimer covering. The average energy imbalance between the left and right parts of the system is depicted in Fig.~\ref{fig:mps_diff}. Due to the large buildup of entanglement in this case, which should be contrasted with the Kibble-Zurek experiments, the MPS method is severely limited and cannot capture the energy diffusion across the system.

\setcounter{manualbibstart}{48}

\end{document}